\documentclass[useAMS,usenatbib]{mn2e}
\usepackage{graphicx}

\title[An H$\alpha$-selected sample of cataclysmic variables -- I.]{An H$\alpha$-selected sample of cataclysmic variables -- I. Observations of newly discovered systems}
\author[M.L. Pretorius and C. Knigge]{Magaretha L. Pretorius\thanks{E-mail: mlp@astro.soton.ac.uk (MLP); christian@astro.soton. ac.uk (CK)} and Christian Knigge\footnotemark[1] \\
School of Physics and Astronomy, University of Southampton, Highfield, Southampton SO17 1BJ, United Kingdom\\}
\begin{document}


\pagerange{\pageref{firstpage}--\pageref{lastpage}} \pubyear{}

\maketitle

\label{firstpage}

\begin{abstract}
Strong selection effects are present in observational samples of cataclysmic variables (CVs), complicating comparisons to theoretical predictions.  The selection criteria used to define most CV samples discriminate heavily against the discovery of short-period, intrinsically faint systems.  The situation can be improved by selecting CVs for the presence of emission lines.  For this reason, we have constructed a homogeneous sample of CVs selected on the basis of H$\alpha$ emission.  We present discovery observations of the 14 CVs and 2 additional CV candidates found in this search.  The orbital periods of 11 of the new CVs were measured; all are above 3~h.  There are two eclipsing systems in the sample, and one in which we observed a quasi-periodic modulation on a $\sim 1\,000\,\mathrm{s}$ time-scale.  We also detect the secondary star in the spectrum of one system, and measure its spectral type.  Several of the new CVs have the spectroscopic appearance of nova-like variables (NLs), and a few display what may be SW Sex star behaviour.  In a companion paper, we discuss the implications of this new sample for CV evolution.
\end{abstract}


\section{Introduction}
Cataclysmic variable stars (CVs) are semi-detached binary stars, consisting of a white dwarf primary accreting from an approximately main-sequence companion (see e.g. \citealt{bible}).  The mass transfer is caused by orbital angular momentum loss, which also drives the secular evolution of CVs.

At long orbital periods ($P_{orb}\ga3\,\mathrm{h}$), angular momentum loss through magnetic braking is thought to dominate over that resulting from gravitational radiation.  According to the `disrupted magnetic braking' model of CV evolution, magnetic braking stops when $P_{orb}$ reaches $\simeq 3\,\mathrm{h}$, the secondary loses contact with its Roche lobe, and gravitational radiation is left as the only angular momentum loss mechanism (e.g. \citealt{RobinsonBarkerCochran81}; \citealt{RappaportVerbuntJoss83}; \citealt{SpruitRitter83}).  The secondary only regains contact with its Roche lobe when gravitational radiation has decreased $P_{orb}$ to $\simeq 2\,\mathrm{h}$.  This model explains the pronounced drop in the number of CVs at $2~\mathrm{h} \la P_{orb}\la3$~h, called the period gap.

Mass loss from the secondary increases its thermal time-scale, so that, for CVs below the period gap, this eventually exceeds the mass-transfer time-scale (even though the mass loss time-scale increases as $M_2$ decreases).  When this happens, the secondary is not able to decrease its radius rapidly enough in response to mass loss, so that the orbital evolution slowly moves back through longer periods (e.g. \citealt{Paczynski81}; \citealt{PaczynskiSienkiewicz81}; \citealt{RappaportJossWebbink82}).  CVs in this final phase of evolution, where $P_{orb}$ is increasing, are referred to as `period bouncers'.  The reversal in the direction of change in $P_{orb}$ causes the period minimum---a sharp cut-off in the $P_{orb}$ distribution of hydrogen-rich CVs at about 76~min.

Magnetic braking is much more efficient than gravitational radiation at removing angular momentum---the resulting mass transfer rates ($\dot{M}$) above the gap are roughly 10 to 100 times larger than below the gap (e.g. \citealt{Patterson84}).  This implies that long-period CVs are intrinsically bright, and also that the long-period phase of the evolution of a CV is very short-lived.  The majority of CVs should therefore be short-period, low-$\dot{M}$ (and therefore intrinsically faint) systems.  In fact, even the time taken to evolve from the bottom of the period gap to the period minimum is short compared to the age of the Galaxy, so that most CVs should be period bouncers.  Population synthesis studies predict the relative sizes of the long-period, short-period, and period bouncer populations (roughly $1:30:70$), as well as the absolute size of the Galactic CV population (e.g. \citealt{Kolb93}; \citealt{HowellRappaportPolitano97}).  However, the predictions of relative sizes of sub-populations, as well as the overall space density of CVs, have been disputed on observational grounds (e.g. \citealt{Patterson84}; \citealt{Patterson98}; \citealt{GansickeHagenEngels02}; \citealt{PretoriusKniggeKolb07}; \citealt{PretoriusNEP}).

A meaningful quantitative comparison between observations and theory is only possible if observational selection effects can be accounted for.  This requires that the criteria for inclusion in the observational sample are well-defined, and preferably homogeneous.  Furthermore, the typical intrinsic brightness of the lowest-$\dot{M}$ CV population that can be probed by a given survey clearly depends on the survey flux limit.  If this limit is too bright, the survey has no sensitivity to any but the intrinsically brightest (and rarest) CVs.  

Several complete, uniformly selected CV samples already exist, e.g. those that resulted from the Palomar Green Survey, the \emph{ROSAT} Bright Survey, and the \emph{ROSAT} North Ecliptic Pole Survey (\citealt{Ringwald93}; \citealt{SchwopeBrunnerBuckley02}; \citealt{PretoriusNEP}).  Ongoing surveys, e.g. the Hamburg Quasar Survey, and the Sloan Digital Sky Survey (\citealt{GansickeHagenEngels02}; \citealt{AungwerojwitGansickeRodriguez-Gil06}; \citealt{sdsscvs1}, 2003, 2004, 2005, 2006, 2007) are in the process of producing similarly well-defined, but much deeper, samples.

All CV samples are affected by a flux-limit\footnote{This may not be important for the discovery of classical novae, but it is for recovering them and measuring periods.}, which already implies a bias against intrinsically faint systems.  But it is in fact very difficult to construct a sample that is reasonably large and deep, and purely flux-limited.  Of the surveys mentioned above, only the very shallow \emph{ROSAT} Bright Survey, and the very small \emph{ROSAT} North Ecliptic Pole Survey do not contain a blue cut, in addition to a flux limit.  The surveys that incorporate a blue selection are biased against low-$\dot{M}$ CVs, since these systems are not only intrinsically faint, but also relatively red.  The single property through which the most CVs have been discovered is large amplitude variability.  It is not easy to quantify the selection effects acting on the CVs discovered in this way, but they certainly also favour intrinsically brighter systems, since low-$\dot{M}$ CVs undergo less frequent outbursts.  \cite{Gansicke05} reviews all surveys that have discovered sizable samples of CVs.

The presence of Balmer emission lines in the spectra of most CVs provides an alternative to the commonly used blue- and variability-based selection techniques.  Selecting CVs for line emission discriminates only against the discovery of the intrinsically rare CVs with very bright, optically thick discs.  In fact, there is a well known (and theoretically expected) anti-correlation between the equivalent widths (EWs) of Balmer emission lines and the luminosity of CVs (\citealt{Patterson84}; \citealt{WithamKniggeGansicke06}; we will take EWs of emission lines as positive throughout).

A few surveys have already exploited the promise of discovering CVs via emission lines.  CVs are selected from the Hamburg Quasar Survey based in part on H$\beta$ emission.  One of the selection criteria for CVs from the Cal\'{a}n-Tololo Survey is Balmer emission lines \citep{TappertAugusteijnMaza04}.  The INT Photometric H$\alpha$ Survey of the Northern Galactic Plane (IPHAS) is currently being used to find CVs (\citealt{WithamKniggeGansicke06}; \citealt{WithamKniggeAungwerojwit07}), and \cite{RogelLuggerCohn06} describe another H$\alpha$-based search for CVs.  H$\alpha$ emission has also been used to identify CV candidates in globular clusters (e.g. \citealt{BailynRubensteinSlavin96}; \citealt{CoolGrindlayCohn95})

The AAO/UKST SuperCOSMOS H$\alpha$ Survey (SHS; \citealt{Parker05}) is currently the best available southern hemisphere resource for identifying H$\alpha$ emission line point sources.  One CV has previously been discovered in this survey, partly on the basis of H$\alpha$ emission \citep{HowellMasonHuber02}.  We have carried out spectroscopic follow-up of sources in this survey with $R<17.5$, selected for H$\alpha$ emission.  This has allowed us to construct a small, homogeneous sample of CVs.

We present discovery spectra of 16 new CV candidates selected from the SHS, as well as time-resolved observations that have confirmed 14 of these systems as CVs, and yielded orbital period measurements for 11 of the new CVs.  The sample will be used to derive constraints on CV evolution theory in a subsequent paper (\citealt{HalphaII}; hereafter Paper II).  

\section{Selection of targets}

\begin{table*}
{\footnotesize
 \centering
  \caption{Coordinates, broad band magnitudes, $R-\mathrm{H}\alpha$ excess, H$\alpha$ equivalent widths, orbital periods, and lower distance limits for the  CVs and CV candidates. 
}
  \label{tab:summary}
  \begin{tabular}{@{}llllllllllllll@{}}
  \hline
Object/coordinates & $B$  & $R$  &$R_2$ &$R_1$ & $I$  & $J$   & $H$   & $K_S$ & $R-\mathrm{H}\alpha$ & $\mathrm{EW}(\mathrm{H}\alpha)/\mathrm{\AA}$ & $P_{orb}/\mathrm{h}$& $d_l/\mathrm{pc}$ & Notes \\
  \hline
073418.56-170626.5 & 17.8 & 16.7 & ---  & 18.3 & 17.4 & 16.23 & 15.93 &$>$14.9& 0.47 & 61(6)         & 3.18542(5)& 320 & 1,2,3\\
074208.23-104932.4 & 16.5 & 15.3 & 15.4 & 16.0 & 14.3 & 14.26 & 13.59 & 13.36 & 1.21 & 63(3), 45(1)  & 5.706(3)  & 490 & 4,5\\
074655.48-093430.5 & 13.8 & 14.0 & 13.6 & 13.4 & 13.3 & 13.47 & 13.29 & 13.19 & 0.46 & 27(1), 23.6(3)& 3.3984(4) & 180 & 6\\
075648.83-124653.5 & 18.3 & 16.0 & 17.0 & 17.0 & 16.2 & 15.39 & 14.73 & 14.61 & 0.61 & 43(7), 47.1(6)& ---       & --- & 7\\
092134.12-593906.7 & 17.3 & 16.8 & 16.7 & 16.9 & 16.7 & 16.49 & 16.20 &$>$15.8& 0.53 & 56(6), 43.7(4)& 3.041(9)  & 530 & 8\\
092751.93-391052.3 & 16.4 & 16.2 & 16.8 & 16.3 & 16.1 & 15.72 & 15.54 &$>$15.7& 0.35 & 15(2), 20.0(3)& 4.1(3)    & 770 & 9\\
094409.36-561711.4 & 17.5 & 16.2 & ---  & 18.4 & 16.7 & 15.72 & 15.40 & 14.96 & 0.99 & 66(4), 67(1)  & 4.506(4)  & 520 & 1,2,3,8\\
102442.03-482642.5 & 17.1 & 16.4 & 17.6 & 16.4 & 16.2 & 15.83 & 15.57 & 15.41 & 0.94 &125(2), 122(1) & 3.673(6)  & 580 & 3\\ 
103135.00-462639.0 & 17.4 & 16.3 & 17.2 & 16.9 & 16.3 & 16.04 & 15.65 & 15.52 & 0.39 & 35(4), 19.4(3)& 3.76(2)   & 630 & 9\\
103959.96-470126.1 & 17.0 & 16.4 & 16.9 & 16.7 & 16.8 & 15.87 & 15.69 & 15.63 & 0.95 & 31(3), 20.2(5)& 3.785(5)  & 670 & 9\\
112921.67-535543.6 & 16.3 & 15.5 & 15.9 & 16.1 & 15.3 & 15.27 & 15.17 & 15.12 & 0.33 & 23(2), 17.4(5)& 3.6851(4) & 510 & 9\\
115927.06-541556.2 & 18.6 & 17.4 & 17.6 & 17.5 & 17.2 & ---   & ---   & ---   & 0.57 & 57(15)        & ---       & --- & 10\\
122105.52-665048.8 & 17.8 & 17.3 & 17.5 & 16.9 & 17.4 & 16.43 &$>$16.3&$>$16.4& 0.58 & 22(3), 66.0(5)& ---       & --- & 7,9,11\\
130559.50-575459.9 & 17.2 & 16.5 & 16.4 & ---  & 16.2 & 15.99 & 15.55 &$>$14.9& 0.59 & 41(4), 58.3(5)& 3.928(13) & 500 & 8\\
163447.70-345423.1 & 16.8 & 16.7 & 16.7 & 17.0 & 16.3 & 15.88 & 15.42 &$>$15.7& 0.61 & 36(5), 34(3)  & ---       & --- & 3,9\\
190039.83-173205.5 & 17.6 & 16.8 & 17.3 & 16.9 & 17.8 & 15.66 & 15.45 & 15.11 & 0.34 & 26(5)         & ---       & --- & 9,10\\
  \hline
  \end{tabular}
\noindent
Notes: (1) Eclipsing system. (2) Possibly an SW Sex star. (3) Large amplitude variability. (4) Secondary star spectral type M$0\pm1$. (5) \emph{ROSAT} source.  (6) Bowen blend detected. (7) Highly variable H$\alpha$ line profile. (8) Strong He\,{\scriptsize II}\,$\lambda$4686 emission. (9) Spectroscopic appearance of a nova-like variable (NL). (10) Classification as a CV not certain. (11) Probable quasi-periodic oscillation (QPO) detected in one light curve.
\hfill
}
\end{table*}

The SHS is a photographic survey performed with the UK Schmidt Telescope (UKST), and scanned by a digitizing machine called SuperCOSMOS.  The survey imaged the southern Galactic plane in $R$ and H$\alpha$, down to a limiting magnitude of $\simeq 20.5$ (see \citealt{Parker05} for a detailed description).  $I$-band photometry from an older UKST survey is included with the H$\alpha$ and $R$ data and was used in our target selection.  

We selected objects that are clear H$\alpha$ excess outliers in the $R-\mathrm{H}\alpha$ vs $R-I$ colour-colour plane for spectroscopic follow-up.  Most of the target sample was restricted to objects brighter than $R=17.0$, but we also considered objects with $17.0 \le R < 17.5$ for one of our identification spectroscopy runs.  Two CV candidates were found in the fainter magnitude bin.  The selection criteria and sample completeness will be discussed in more detail in Paper II.  

\begin{figure*}
 $\begin{array}{c@{\hspace{0.1cm}}c@{\hspace{0.1cm}}c@{\hspace{0.1cm}}c}
 \includegraphics[width=43mm]{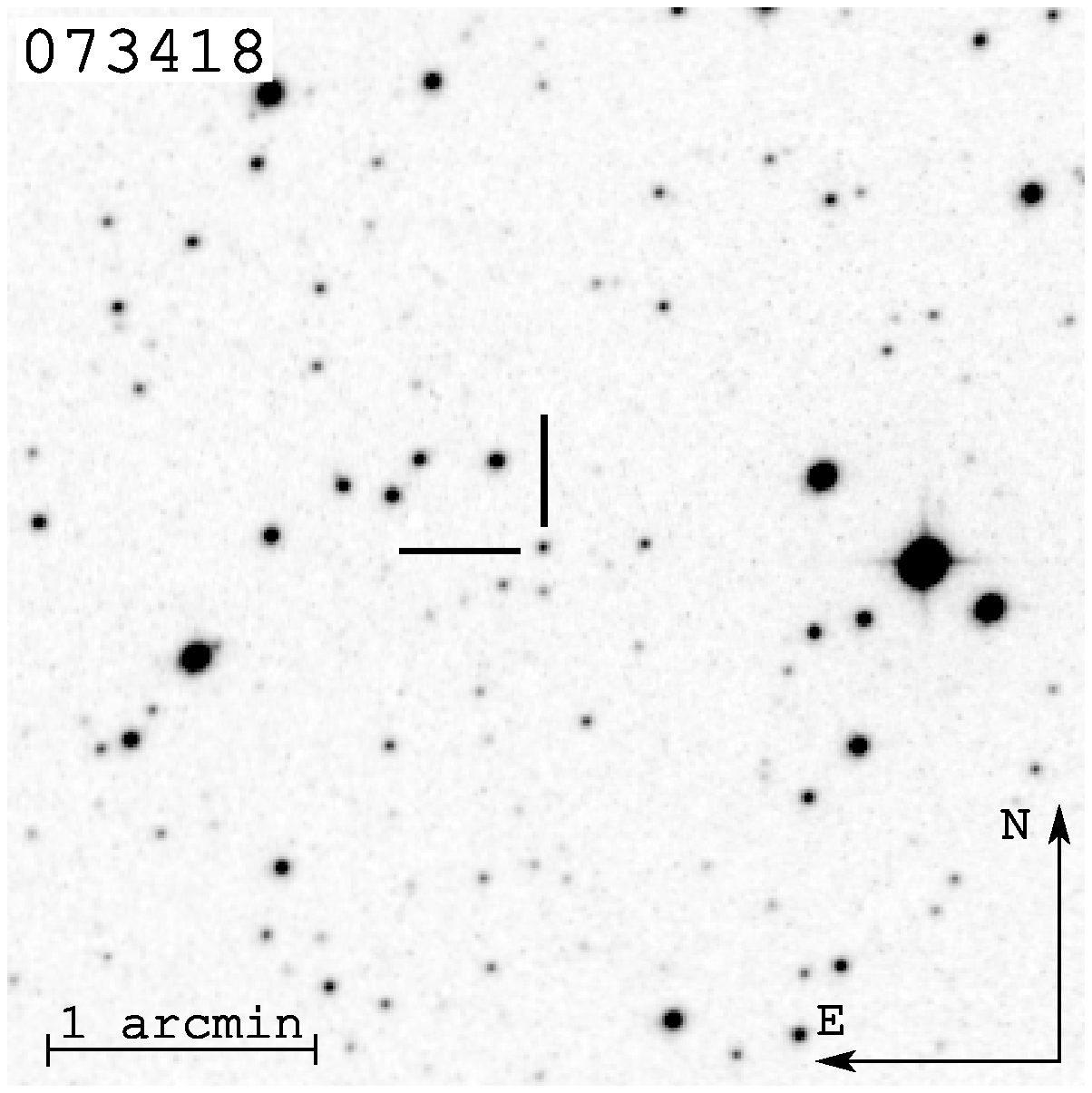} &
 \includegraphics[width=43mm]{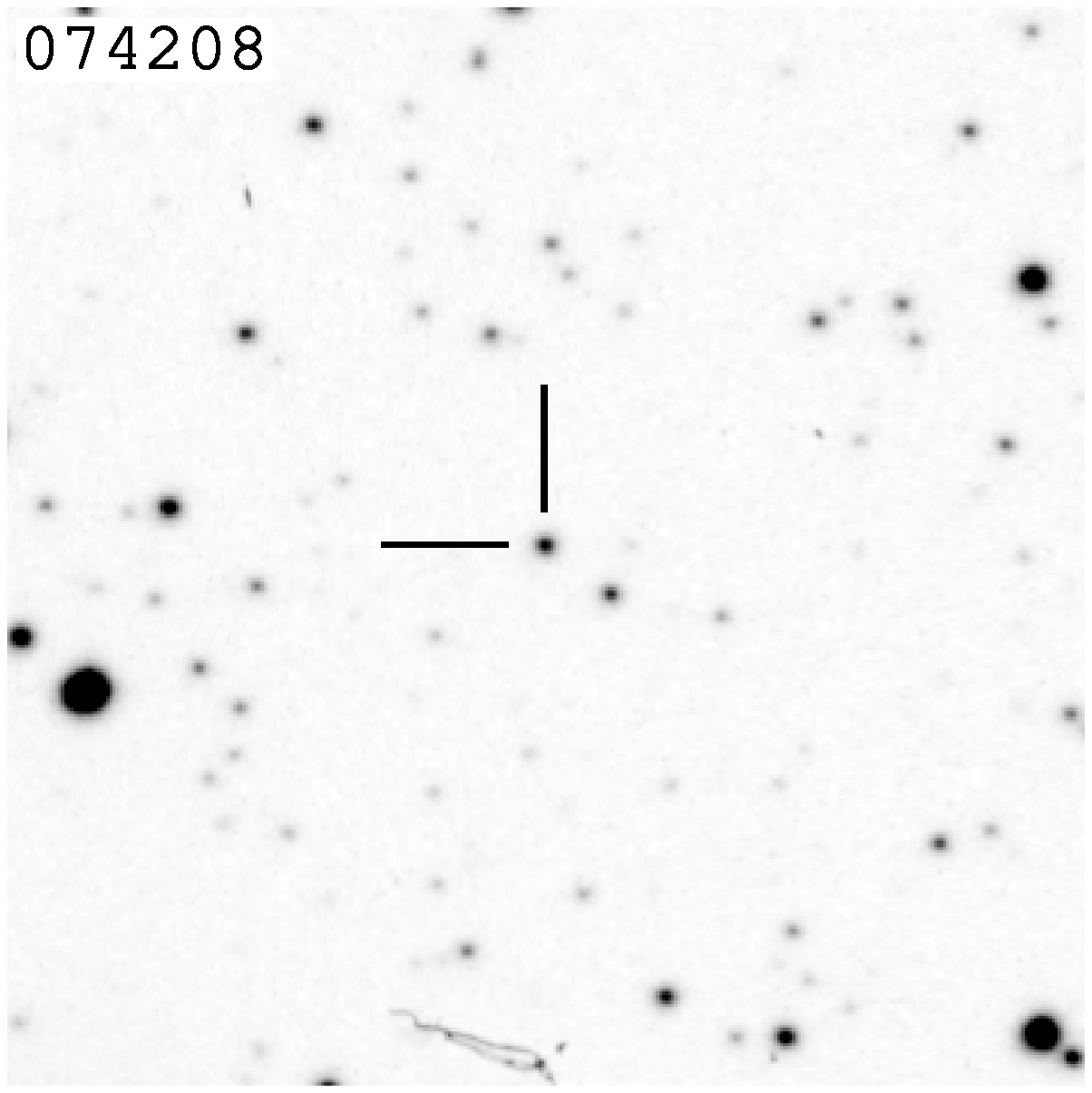} &
 \includegraphics[width=43mm]{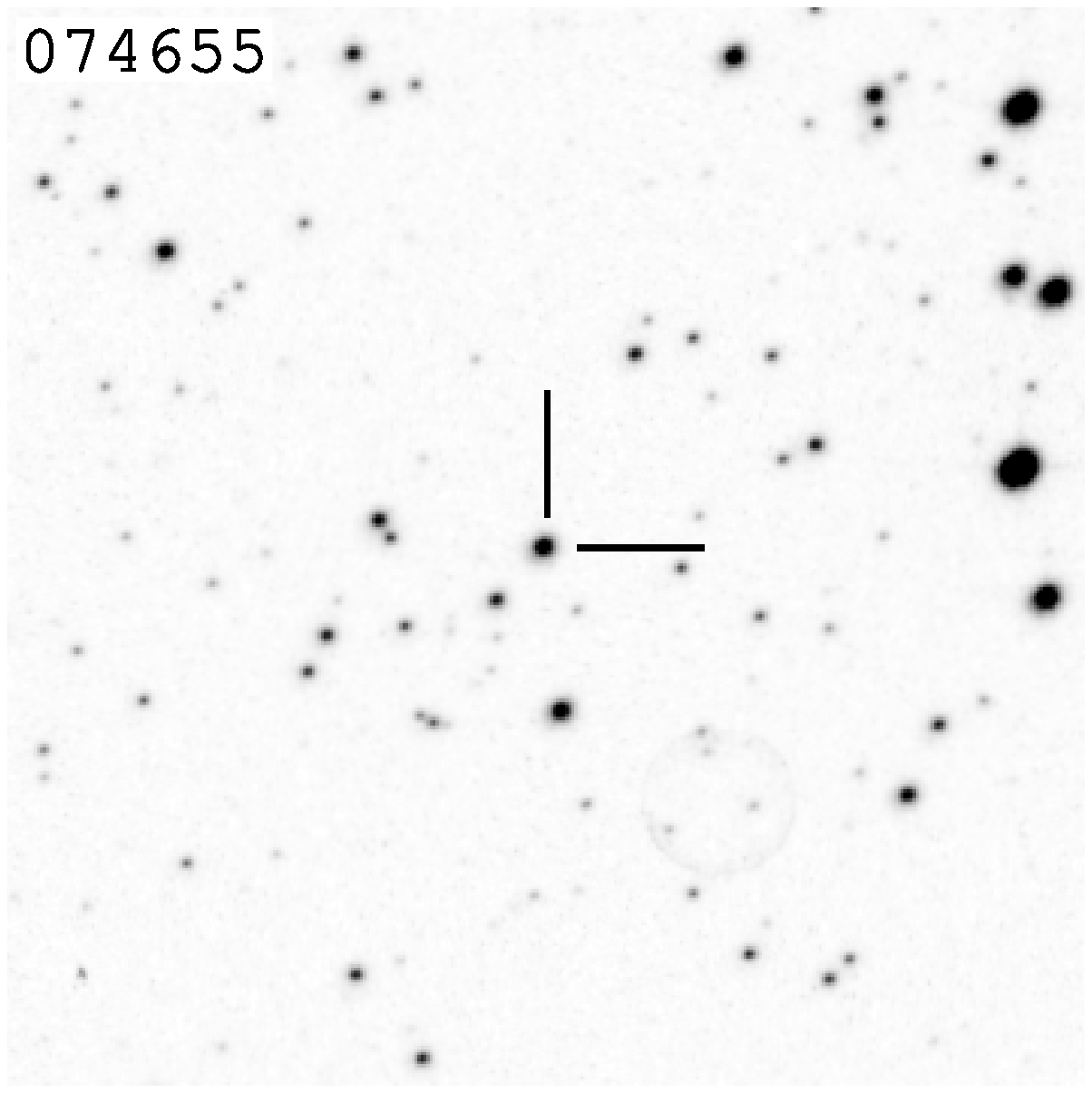} &
 \includegraphics[width=43mm]{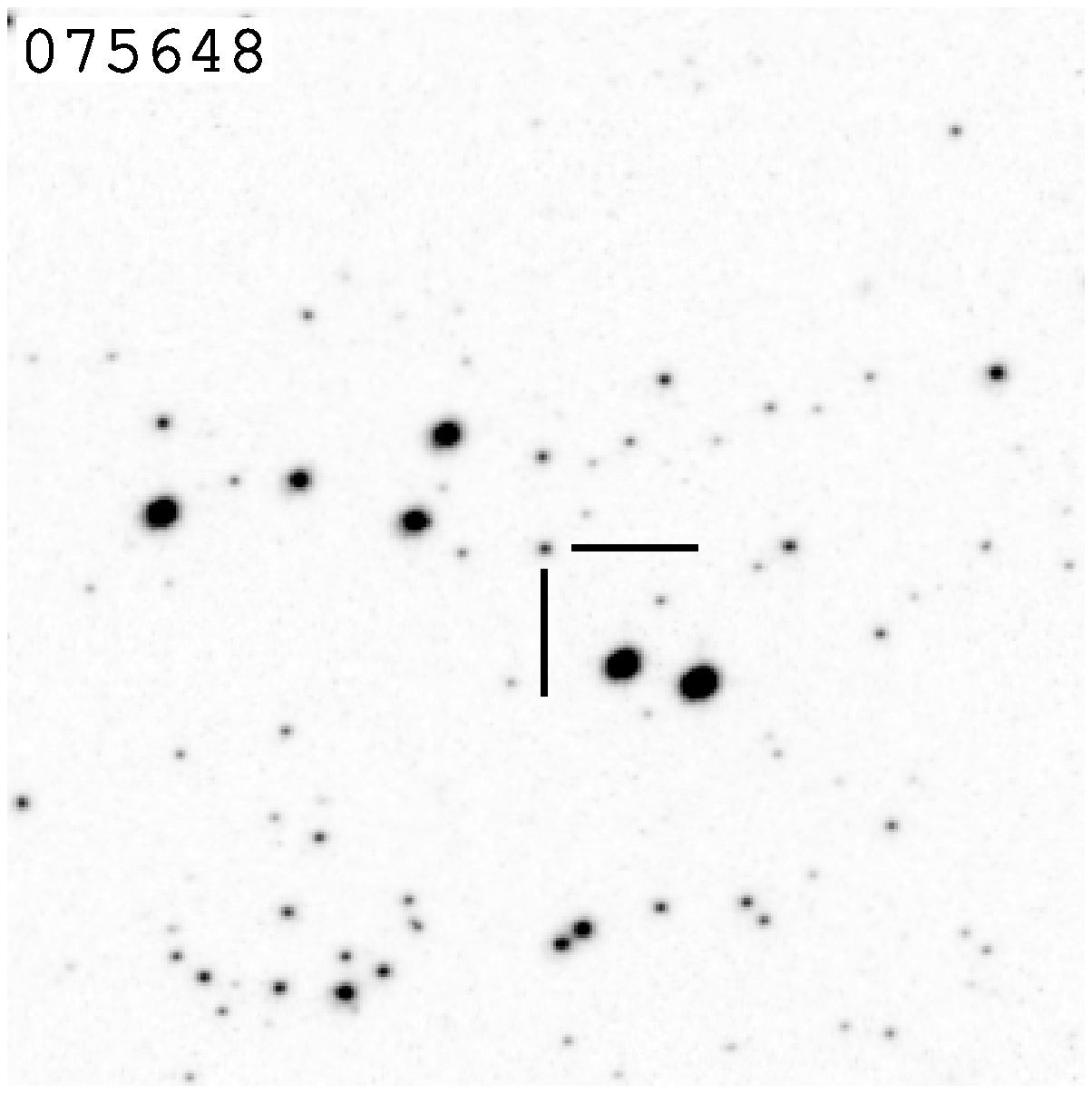} \\
 \includegraphics[width=43mm]{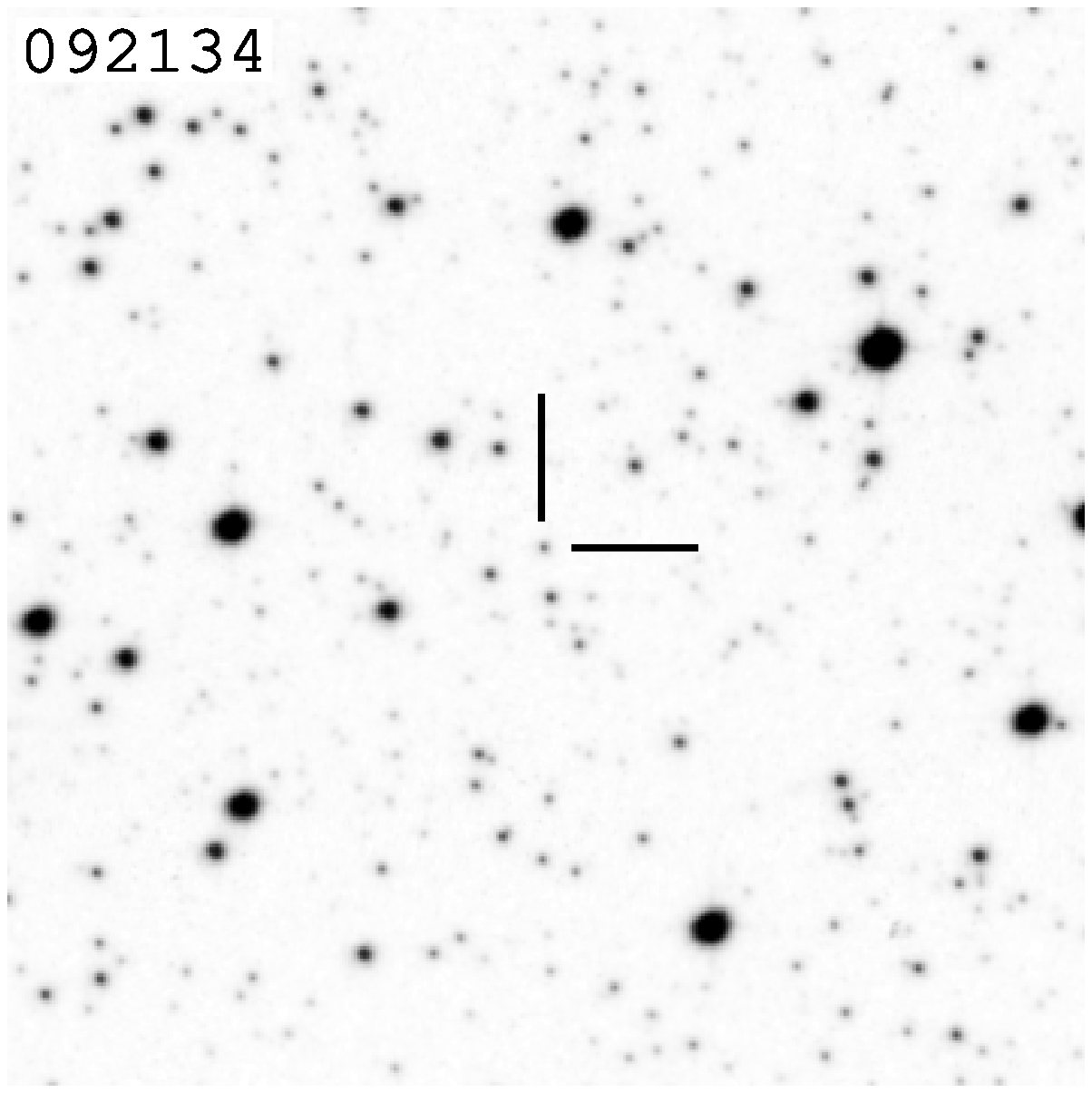} &
 \includegraphics[width=43mm]{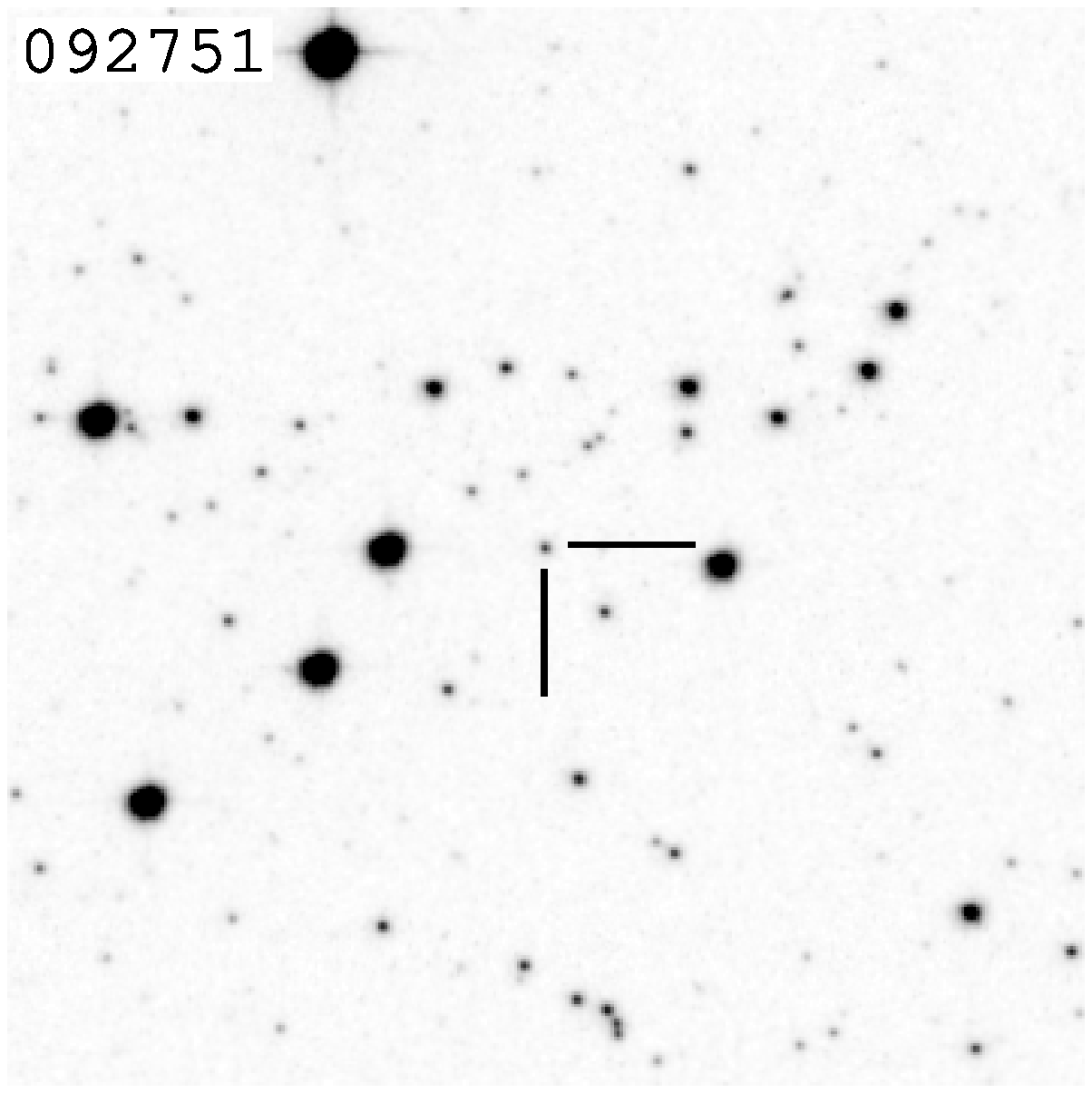} &
 \includegraphics[width=43mm]{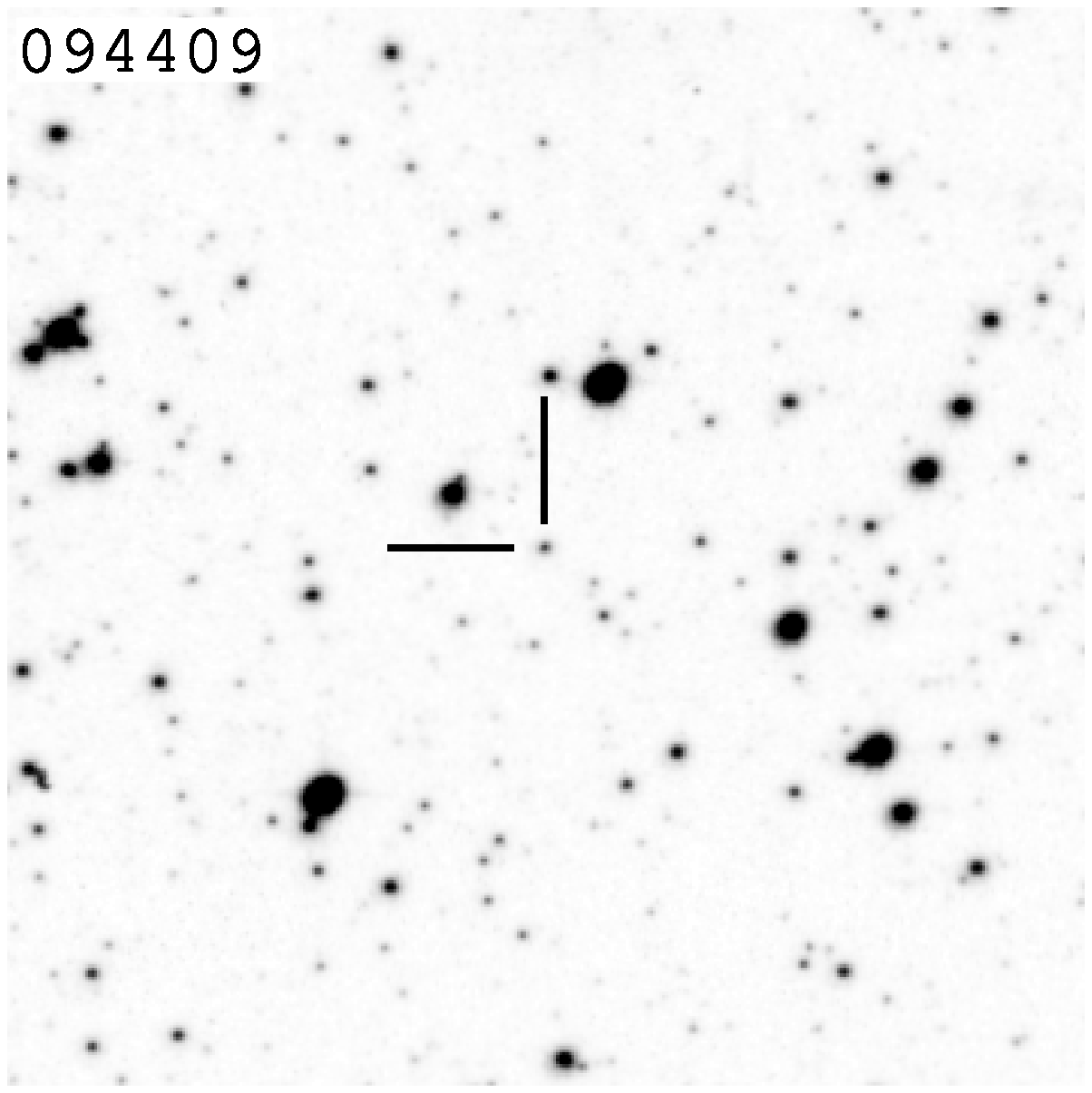} &
 \includegraphics[width=43mm]{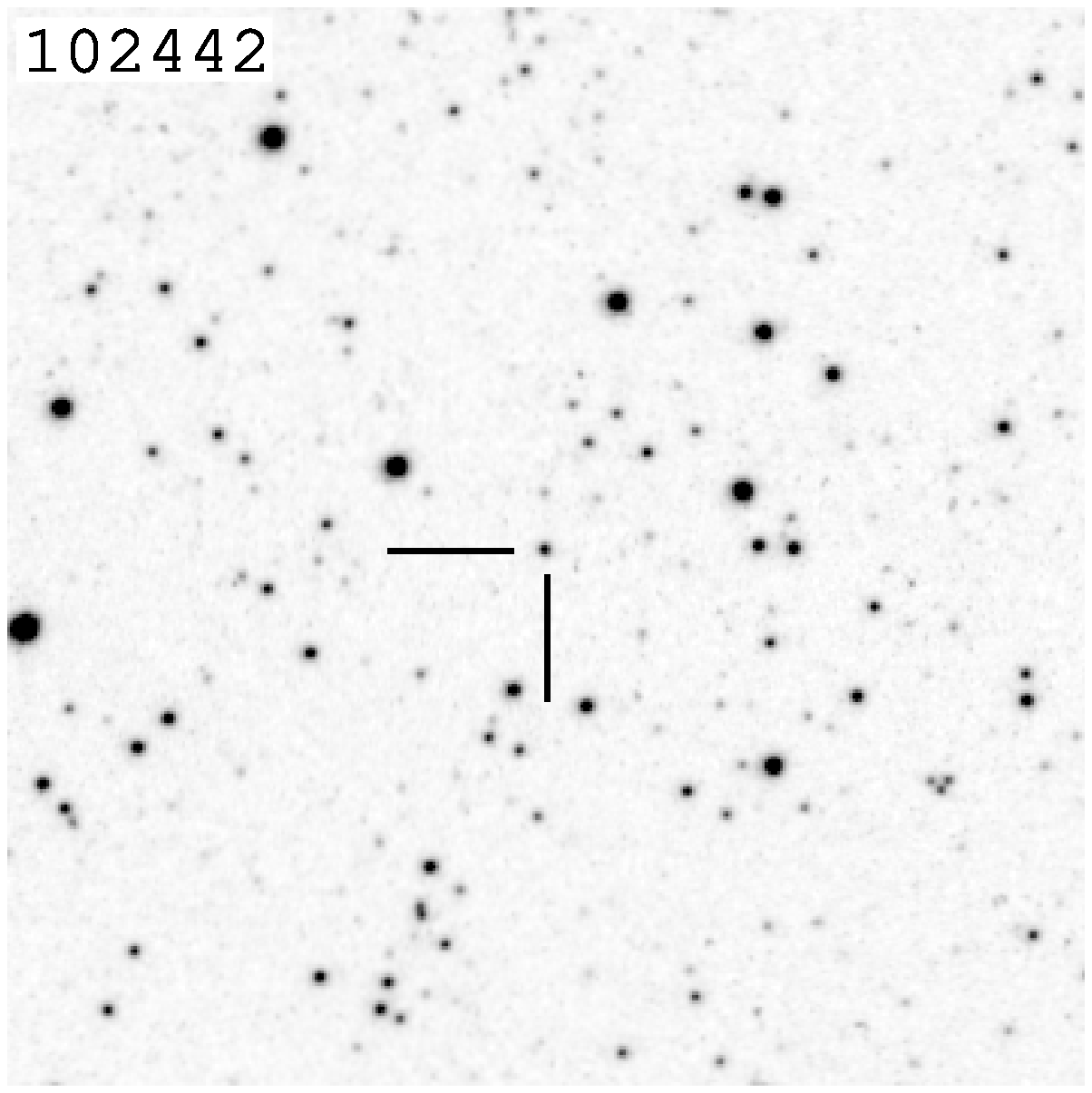} \\
 \includegraphics[width=43mm]{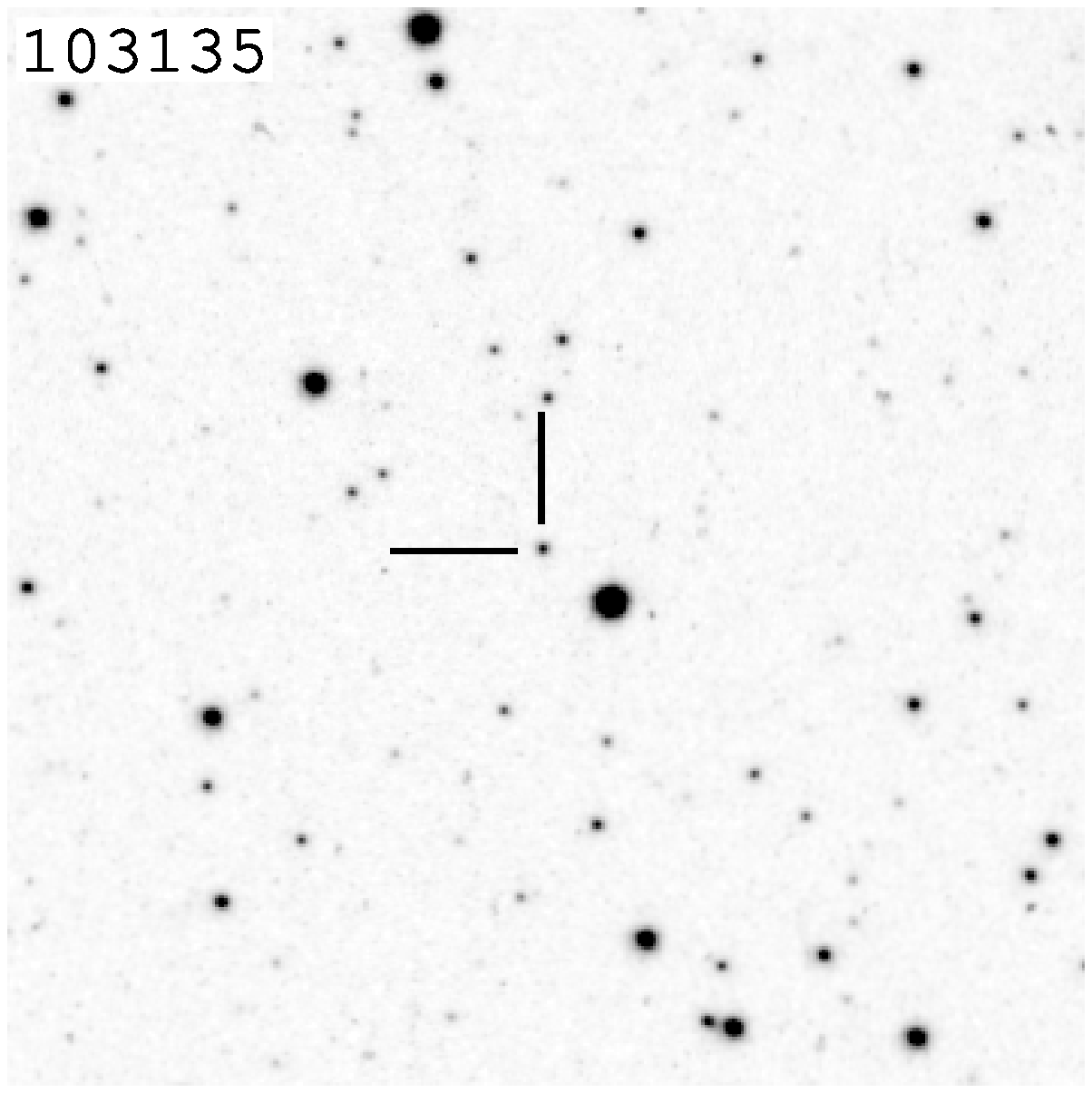} &
 \includegraphics[width=43mm]{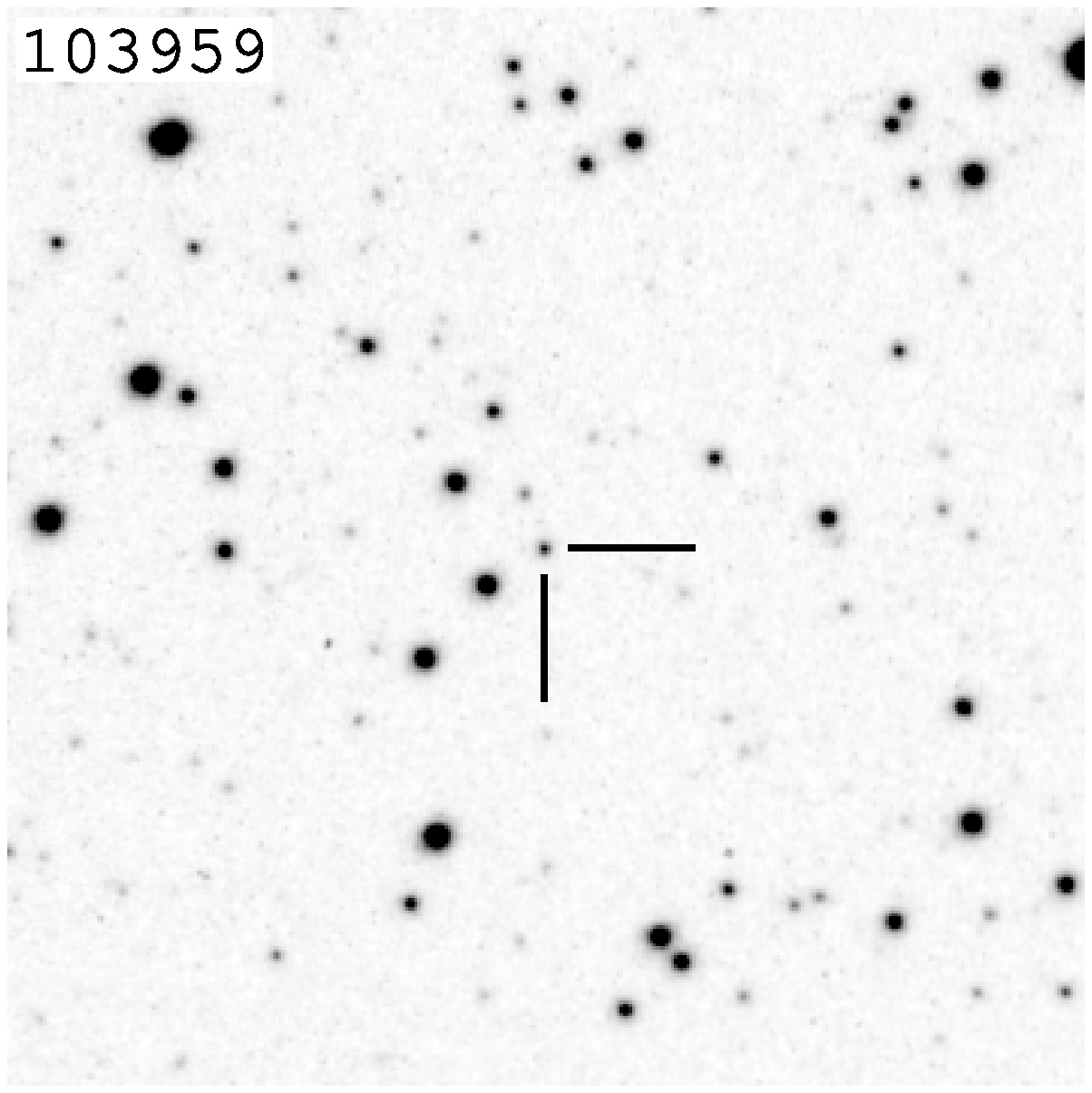} &
 \includegraphics[width=43mm]{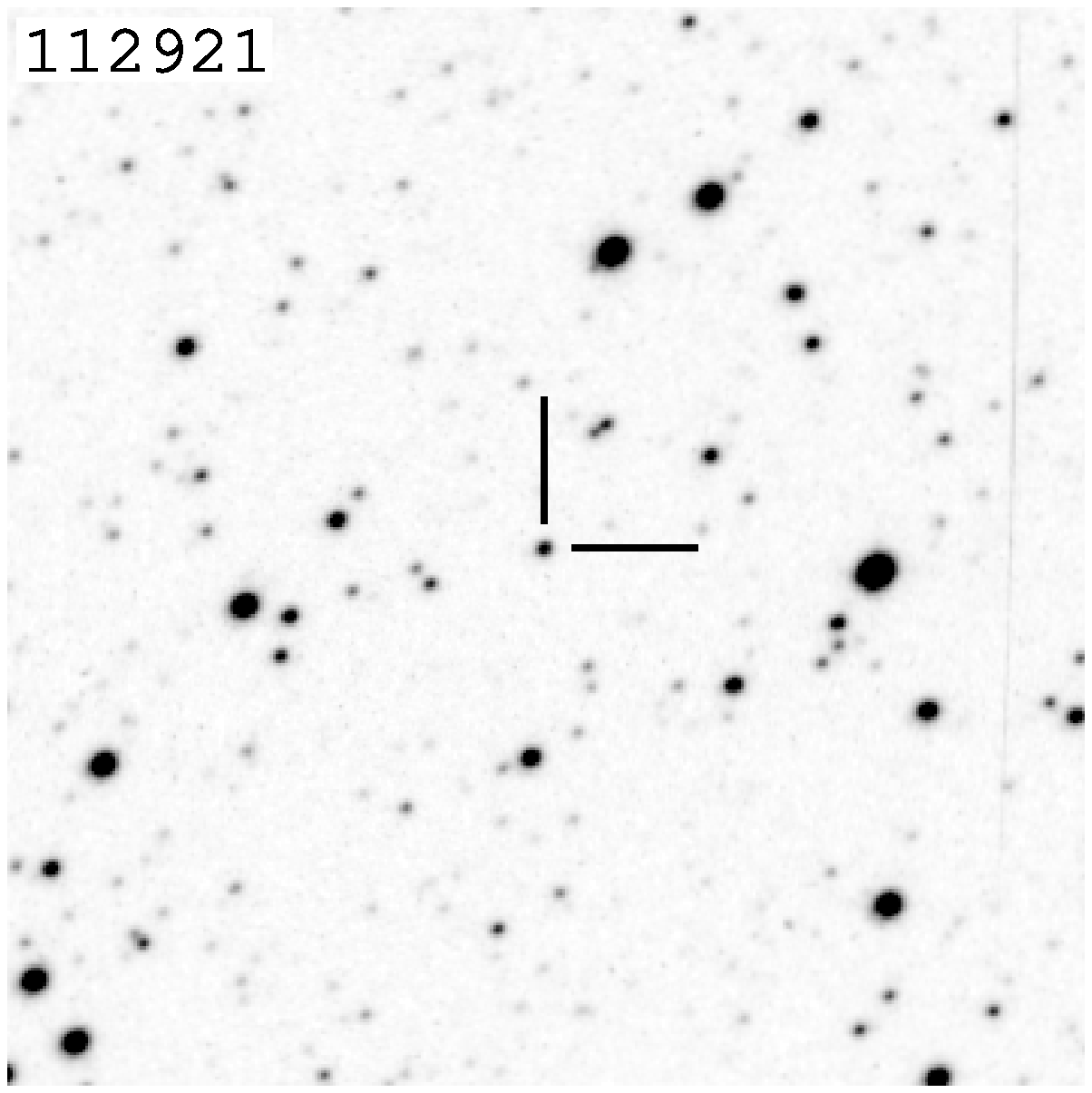} &
 \includegraphics[width=43mm]{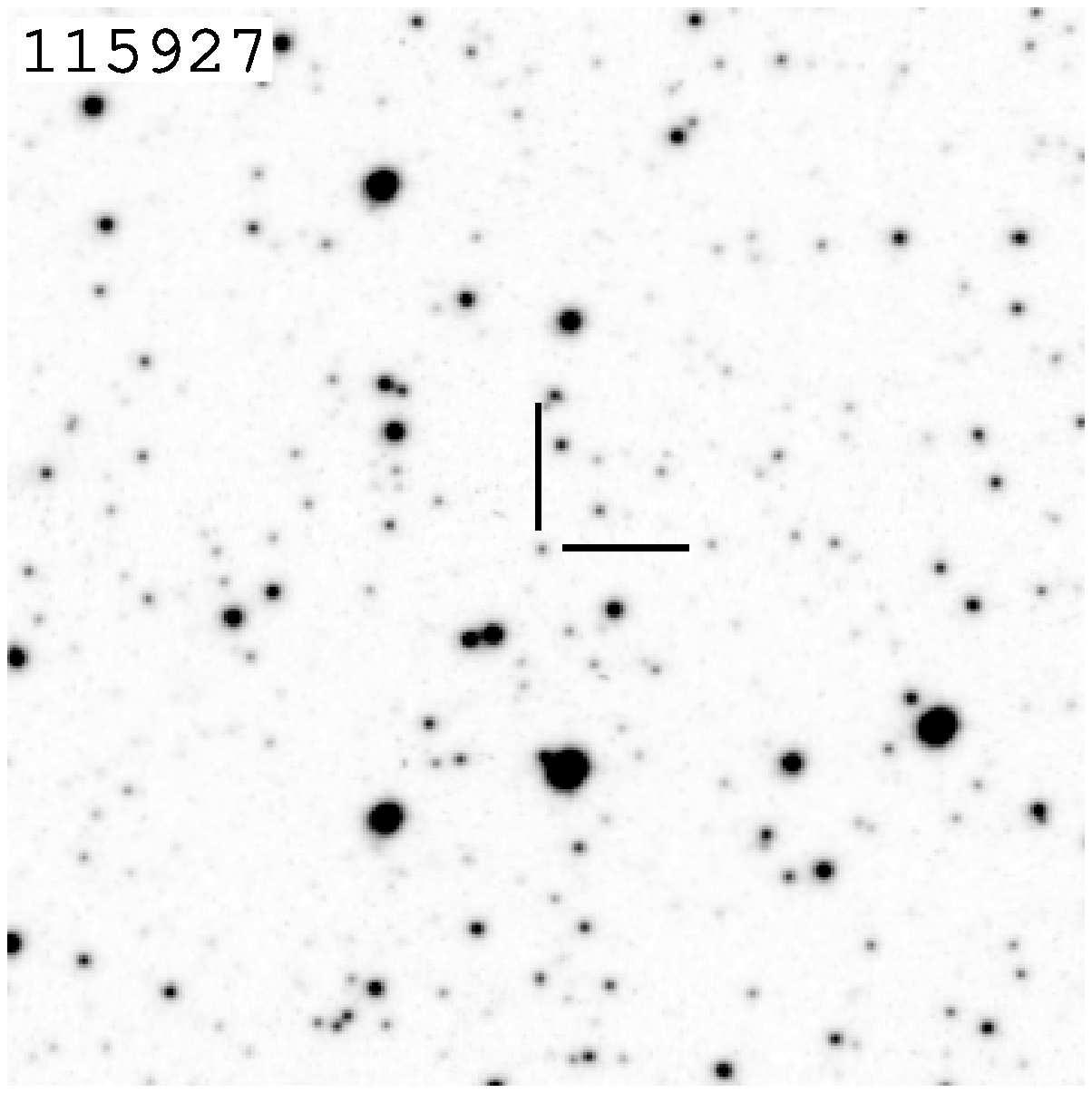} \\
 \includegraphics[width=43mm]{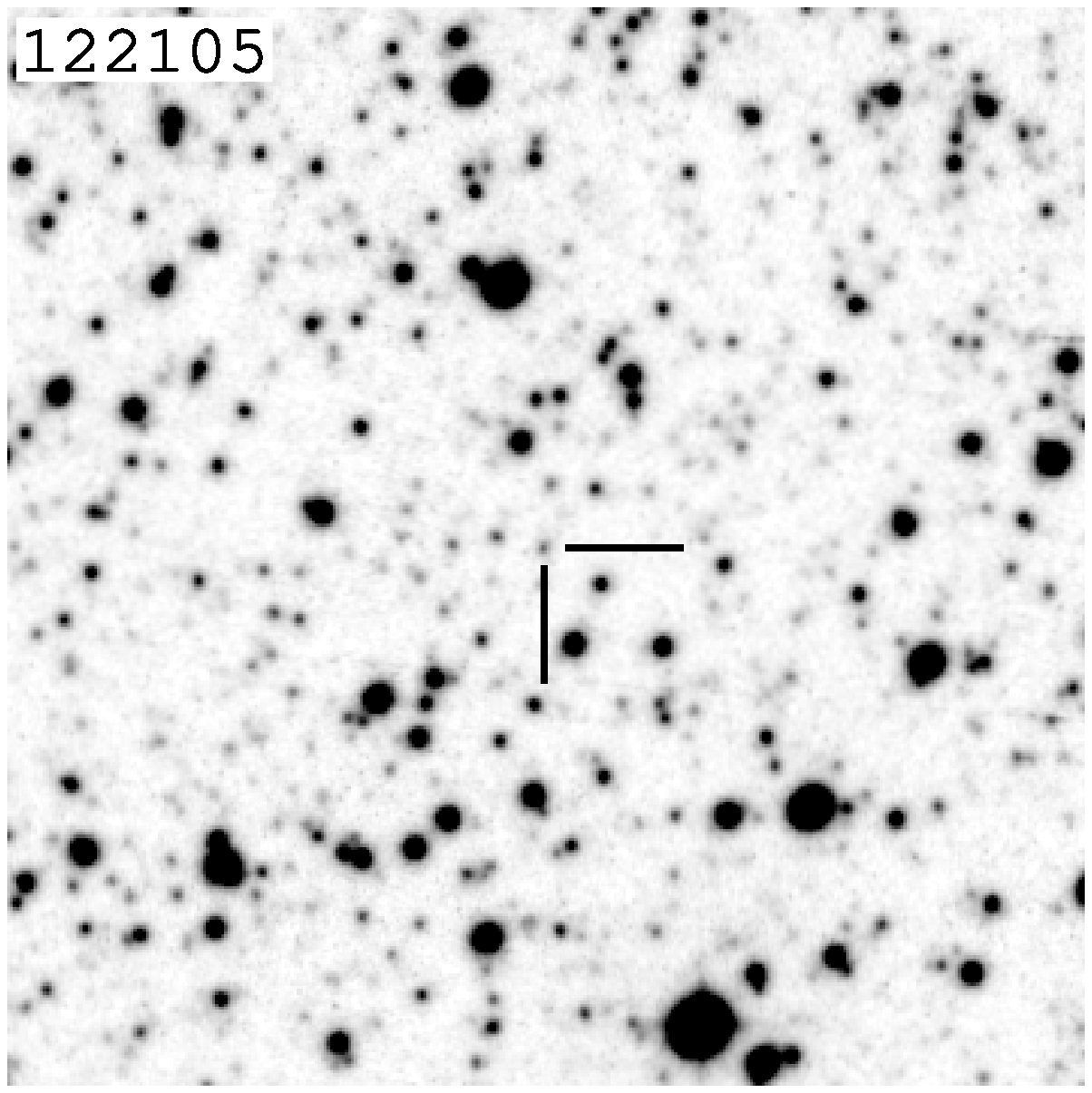} &
 \includegraphics[width=43mm]{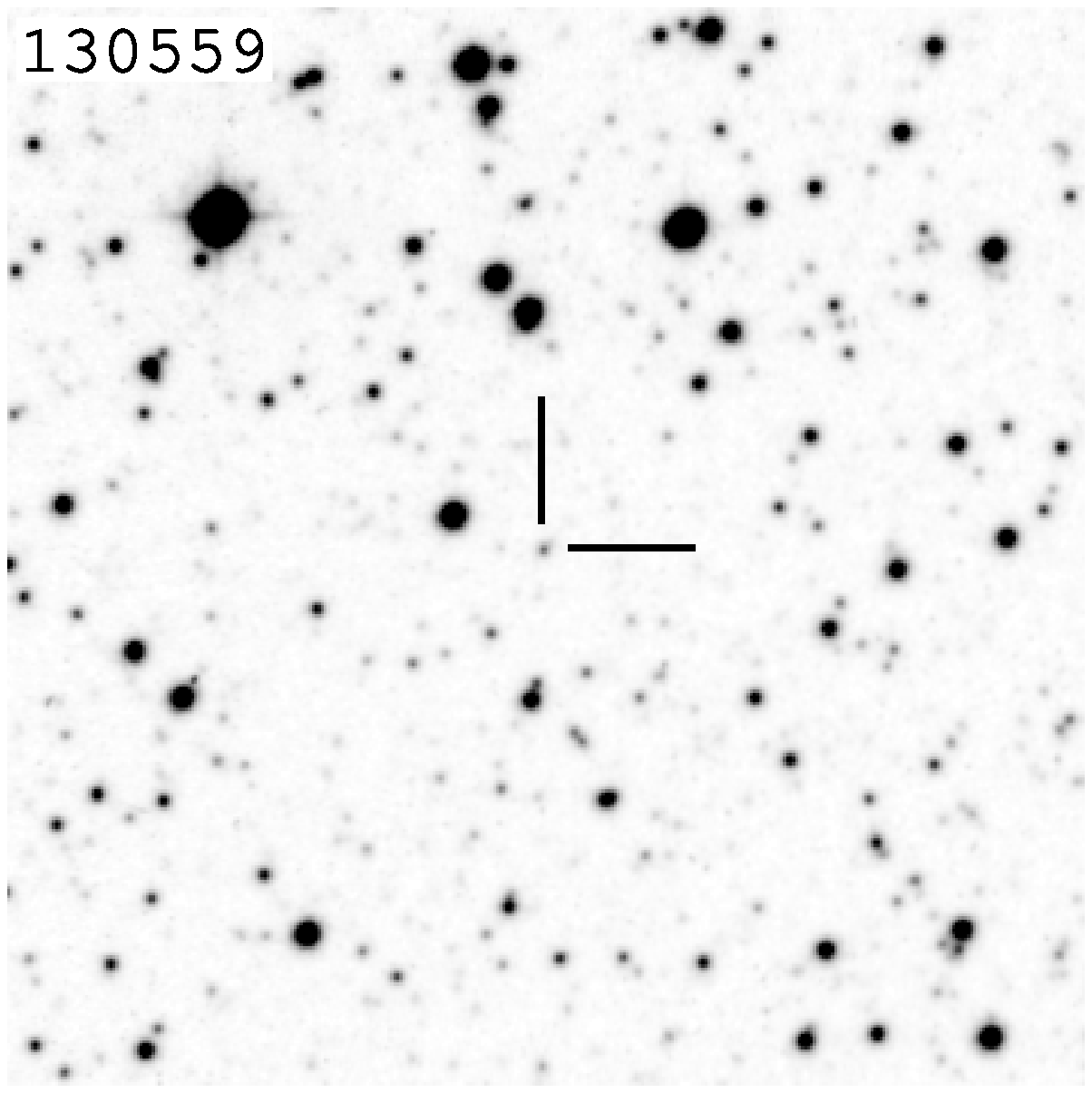} &
 \includegraphics[width=43mm]{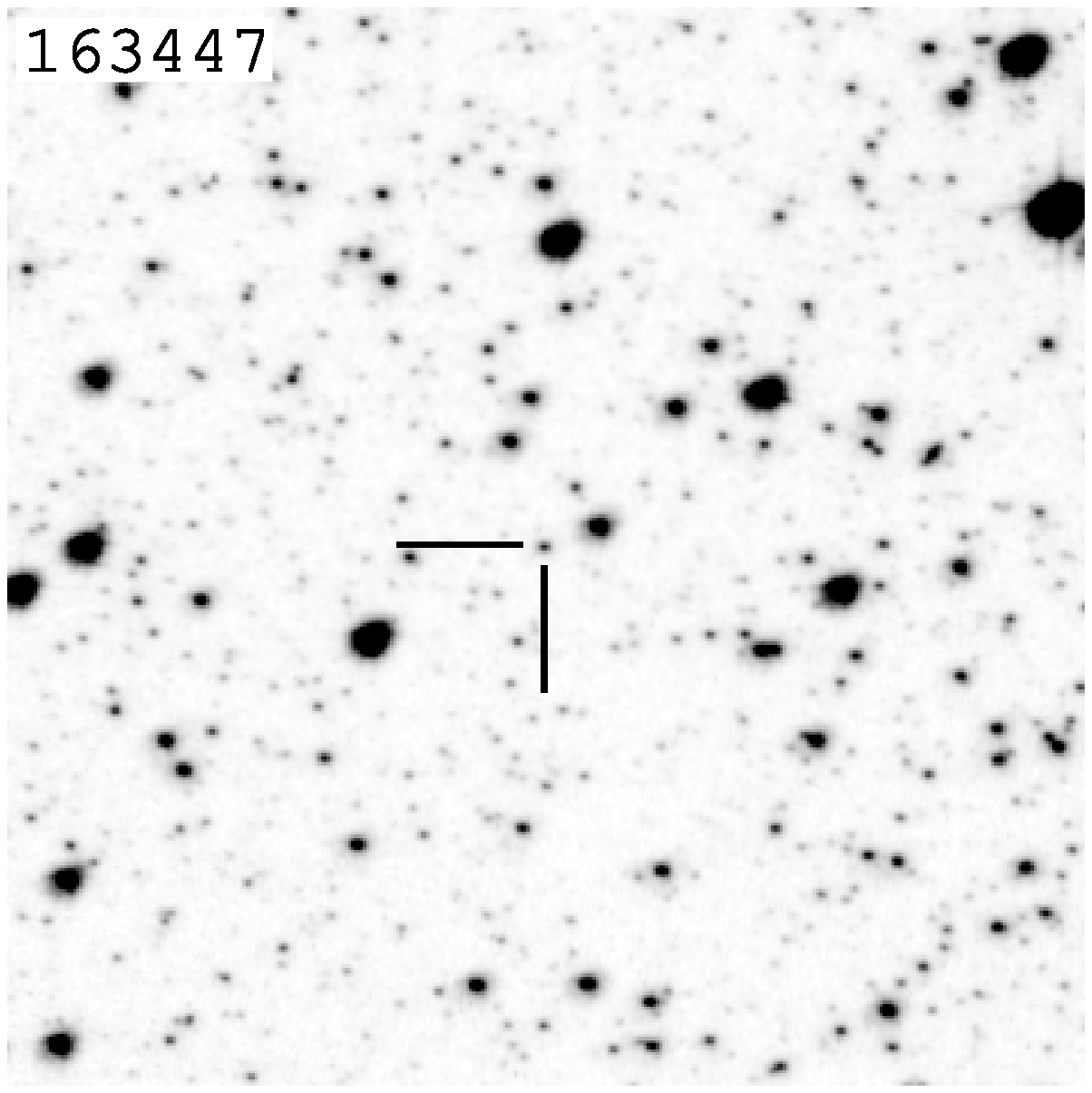} &
 \includegraphics[width=43mm]{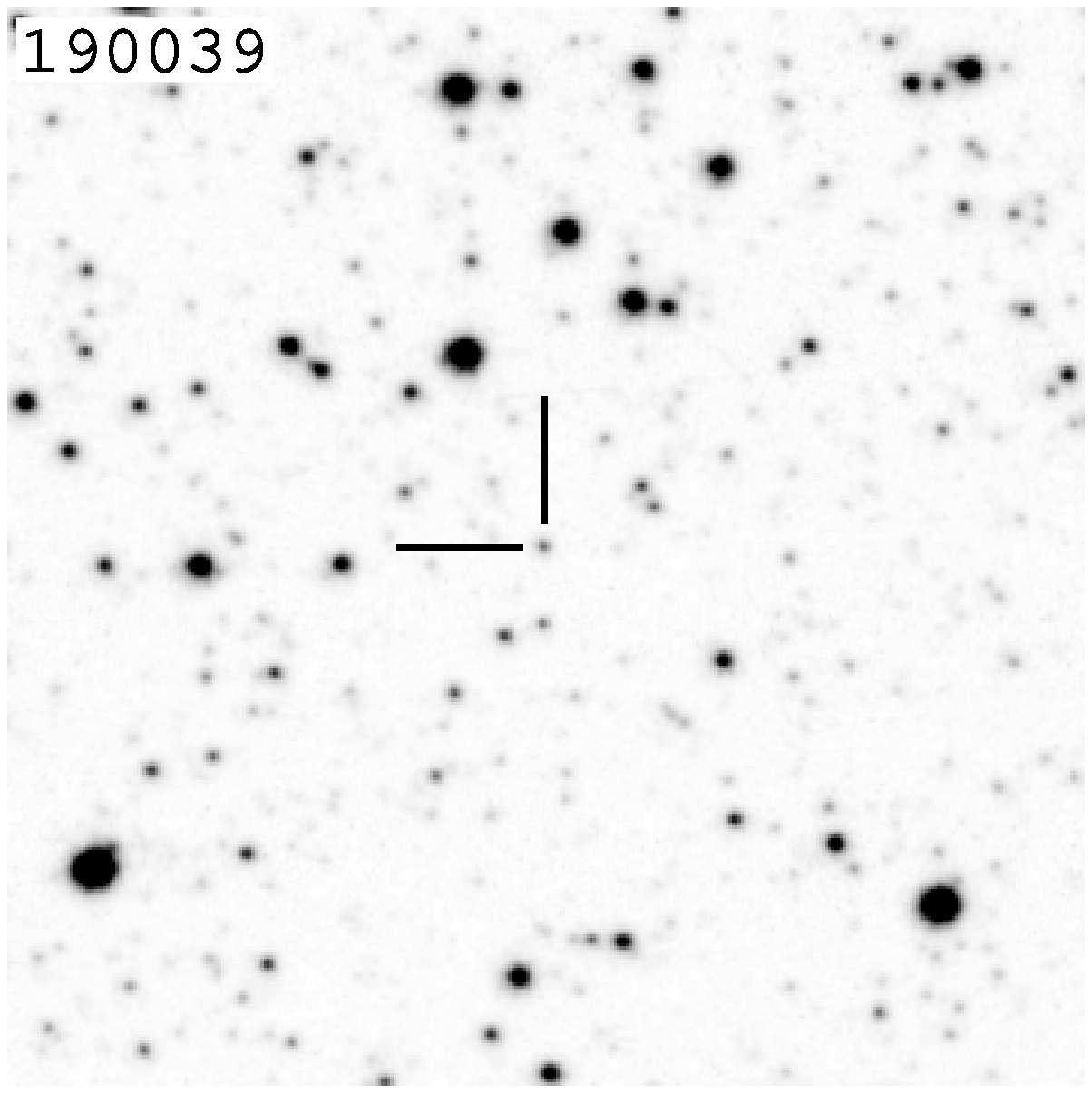} 
 \end{array}$
 \caption {$4' \times 4'$ finding charts of the new CVs and CV candidates, made using UKST $R$-band plates digitized by SuperCOSMOS.  The charts are labelled with the right ascension of the sources.  North is at the top and east to the left in all images.  
}
 \label{fig:findingcharts}
\end{figure*}

Fig.~\ref{fig:findingcharts} gives finding charts for the new CVs and CV candidates turned up by the search, and Table~\ref{tab:summary} lists J2000 coordinates, broad band magnitudes, H$\alpha$ excesses and equivalent widths, orbital periods, and lower limits on distances.  We will name objects using their right ascension as `H$\alpha$hhmmss'.  Where two values of $\mathrm{EW}(\mathrm{H}\alpha)$ are given, the first is obtained from the identification spectrum, and the second from the average of higher resolution spectra of that system.  Errors in the equivalent widths were estimated using the method described by \cite{HowarthPhillips86}.  Near-infrared magnitudes are from the Two Micron All Sky Survey (2MASS; \citealt{2mass})\footnote{H$\alpha$092751 was not detected in $K_S$ by 2MASS; the 95\% confidence lower limit on $K_S$ is given as 16.65.  This implies either very unusual near-IR colours for the source, or variability between the different IR images (which were taken on the same night).  However, the number-magnitude count of sources detected in a 5 arc min radius around the position of H$\alpha$092751 indicates that the magnitude limit of the $K_S$-band image is about 15.7.  We will adopt this more conservative limit.}.  The optical magnitudes listed in Table~\ref{tab:summary} are from SuperCOSMOS scans of the UKST Blue Southern and Equatorial Surveys ($B$), the UKST Red Southern and Equatorial Surveys ($R_2$), the ESO Schmidt Telescope Red Southern Survey ($\delta < -17.5^\circ$) or (for $-17.5^\circ < \delta < +2.5^\circ$) the Palomar-I Oschin Schmidt Telescope (POSS-I) Red Southern Extension ($R_1$), and the UKST near Infrared Southern Survey ($I$).  The orbital period measurements are discussed in Section~\ref{sec:results}, and lower limits on the distances to our sources ($d_l$) are derived in Section~\ref{sec:distances}.

\section{Observations}
We obtained follow-up observations at the South African Astronomical Observatory (SAAO), the European Southern Observatory (ESO), and the Cerro Tololo Inter-American Observatory (CTIO).  We initially took identification spectra of 460 objects (out of a total of 507 targets; see Paper~II).  Time-resolved observations of the objects that we classified as CV candidates were obtained later.

\subsection{Identification spectroscopy}
Medium-resolution identification spectra were taken with the Grating Spectrograph on the SAAO 1.9-m telescope.  The no. 7 grating was used in combination with a slit width of $1\farcs8$, yielding spectral resolution of $\simeq 5\,\mathrm{\AA}$\ over the wavelength range 3\,700 to 7\,200~\AA.  Each object spectrum was bracketed with arc lamp exposures (taken with the telescope in the same position) to provide wavelength calibration.  The flux calibration was achieved by observing spectrophotometric standard stars from \cite{StoneBaldwin83} and \cite{Hamuy94} on all except the nights of worst transparency.  Note, however, that many of the identification spectra were obtained under non-photometric conditions; the absolute flux calibrations of some of the spectra shown below are therefore not reliable.  The data were reduced using standard procedures in {\sc iraf}\footnote{{\sc iraf} is distributed by the National Optical Observatories.}, including optimal extraction \citep{Horne86}.

\begin{figure*}
 \includegraphics[width=178mm]{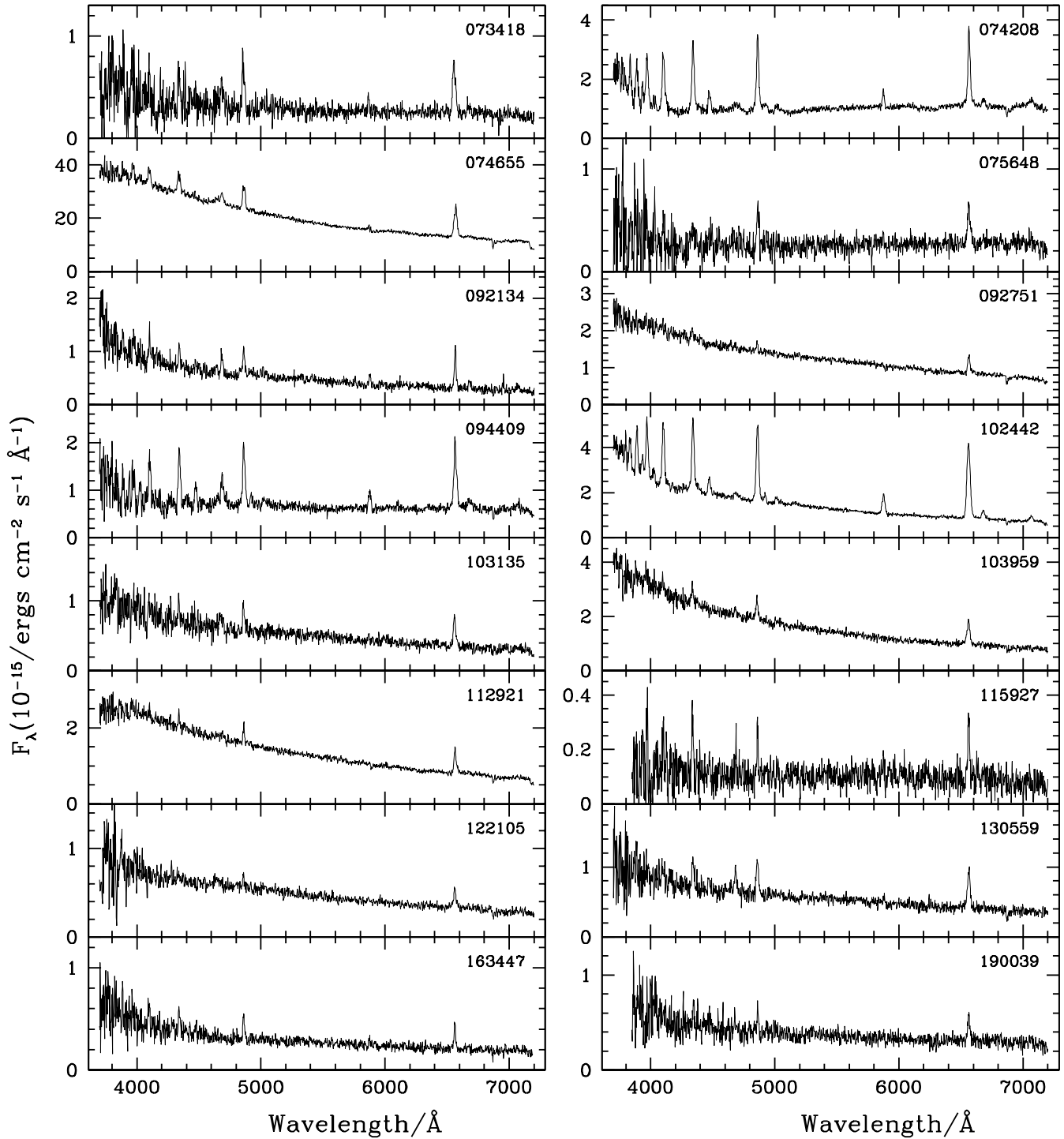}
 \caption{Identification spectra of the newly discovered CVs and CV candidates, obtained with the SAAO 1.9-m telescope.  The panels are labelled with the right ascensions of the sources.  All objects show the broad Balmer emission lines commonly seen in CVs.  He\,{\scriptsize I} lines are detected in most, and He\,{\scriptsize II}\,$\lambda$4686 in some cases.  For the systems where we list two observations in Table~\ref{tab:log_id} (H$\alpha$092751, H$\alpha$122105, and H$\alpha$130559) the spectrum shown here is the average of the two.
}
 \label{fig:idspectra}
\end{figure*}

Further details of the discovery observations are listed in Table~\ref{tab:log_id}.  Spectra of the 16 objects identified as CV candidates are shown in Fig.~\ref{fig:idspectra}.  Our classification of an object as a CV candidate relies mainly on the detection of broad Balmer emission lines.  

\begin{table}
 \centering
  \caption{The discovery observations. Dates are for the start of the night, and $t_{int}$ is integration time.}
  \label{tab:log_id}
  \begin{tabular}{@{}llll@{}}
  \hline
Object          & Date        & HJD $2453000.0 +$ & $t_{int}$/s \\
  \hline
H$\alpha$073418 & 2006 Mar 13 & 808.27607324      & 1200 \\
H$\alpha$074208 & 2006 Mar  4 & 799.37595528      & 1100 \\
H$\alpha$074655 & 2006 Feb 28 & 795.34335764      &  900 \\
H$\alpha$075648 & 2006 Apr  5 & 831.26266972      & 1200 \\
H$\alpha$092134 & 2006 Mar  4 & 799.40790090      & 1300 \\
H$\alpha$092751 & 2006 Mar  4 & 799.42938229      & 1200 \\
                & 2006 Mar  5 & 800.40408050      & 1200 \\
H$\alpha$094409 & 2006 Mar  5 & 800.43760606      & 1200 \\
H$\alpha$102442 & 2006 Mar  7 & 802.43638892      & 1000 \\
H$\alpha$103135 & 2006 Mar 11 & 806.42886098      & 1300 \\
H$\alpha$103959 & 2006 Mar  3 & 798.45787022      & 1200 \\
H$\alpha$112921 & 2006 Mar  7 & 802.48228168      & 1100 \\
H$\alpha$115927 & 2005 Jun  6 & 528.25724689      & 1300 \\
H$\alpha$122105 & 2005 Jun  2 & 524.27931099      & 1300 \\
                & 2006 Mar  5 & 800.52630675      & 1200 \\
H$\alpha$130559 & 2006 Mar  3 & 798.54565116      & 1300 \\
                & 2006 Mar  3 & 798.56223782      & 1300 \\
H$\alpha$163447 & 2006 Apr  4 & 830.54425480      & 1200 \\
H$\alpha$190039 & 2005 Jul 26 & 578.50057109      & 1800 \\
  \hline
  \end{tabular}
\end{table}

\subsection{Time-resolved spectroscopy}
In order to confirm the nature of our CV candidates and measure the orbital periods, we obtained medium-resolution, time-resolved spectroscopy of 13 of the systems.  The brighter objects were observed with the SAAO 1.9-m telescope and the Grating Spectrograph, using grating no. 6 and a slit width of $1\farcs5$.  These spectra have $\simeq 2\,\mathrm{\AA}$ resolution over the wavelength range 5\,220 to 6\,960~\AA.  We obtained additional data for 3 of the SAAO targets with the Cassegrain spectrograph on the CTIO 1.5-m telescope.  The spectrograph was equipped with grating no. 47, which produced a spectral resolution of $\simeq 3.1\,\mathrm{\AA}$, and covered wavelengths between roughly 5\,740 and 6\,850~\AA.  Our fainter targets were observed with the ESO New Technology Telescope (NTT) and the ESO Multi-Mode Instrument (EMMI).  We used the Red Medium-Dispersion (REMD) mode with grating no. 7, in combination with a $1''$ slit.  These spectra cover the wavelength range 5\,550 to 7\,035~\AA\ at $\simeq 2.3\,\mathrm{\AA}$ resolution.  The observing log for the time-resolved spectroscopy is given in Table~\ref{tab:tr_spect}.  

Regular arc lamp exposures were taken to maintain an accurate wavelength calibration, and spectrophotometric standards were observed on each night to allow for flux calibration (as with the identification spectroscopy, many of these observations were obtained under conditions of poor transparency and seeing, but, since the purpose of the observations is to measure orbital periods, the flux calibration is not important).  The time-resolved spectra were also extracted using {\sc iraf}.  The wavelength calibration for each object spectrum was performed by interpolating the dispersion solutions (from fits to the positions of the arc lines) of the two arc lamp spectra taken before and after that object spectrum, and nearest to it in time.

\begin{table*}
 \centering
  \caption{Log of the time-resolved spectroscopy.  Dates are for the start of the night, and $t_{int}$ is the integration time.}
  \label{tab:tr_spect}
  \begin{tabular}{@{}llllll@{}}
  \hline
Object          & Date        & HJD $2450000.0 +$ & $t_{int}$/s & no. of spectra & telescope \\
  \hline
H$\alpha$074208 & 2007 Jan 23 & 4124.57268179  &   600     & 36 & CTIO 1.5-m \\
                & 2007 Jan 31 & 4134.29443787  &   900,800 & 21 & SAAO 1.9-m \\
                & 2007 Feb  2 & 4134.29443787  &   900     & 12 & SAAO 1.9-m \\
                & 2007 Feb  3 & 4135.28935988  &   900     & 22 & SAAO 1.9-m \\
H$\alpha$074655 & 2007 Feb  7 & 4139.34209354  &   900,600 & 31 & SAAO 1.9-m \\
                & 2007 Feb  8 & 4140.27308756  &   600     & 18 & SAAO 1.9-m \\
                & 2007 Feb  9 & 4141.27007383  &   600     & 25 & SAAO 1.9-m \\
                & 2007 Jan 25 & 4126.58152204  &   600     & 27 & CTIO 1.5-m \\
H$\alpha$075648 & 2007 Mar  4 & 4164.65852881  &   800     &  9 & ESO NTT    \\
                & 2007 Mar  5 & 4165.54372639  &   800     & 21 & ESO NTT    \\
H$\alpha$092134 & 2007 Mar  7 & 4167.52424914  &   700     & 16 & ESO NTT    \\
                & 2007 Mar  8 & 4168.52563157  &   700     & 13 & ESO NTT    \\
H$\alpha$092751 & 2007 Mar  6 & 4166.52376369  &   600     & 24 & ESO NTT    \\
H$\alpha$094409 & 2007 Feb  6 & 4138.30240186  &  1000     & 22 & SAAO 1.9-m \\
                & 2007 Feb 12 & 4144.29204597  &  1000     & 18 & SAAO 1.9-m \\
                & 2007 Feb 13 & 4145.29538755  &  1000     & 20 & SAAO 1.9-m \\
H$\alpha$102442 & 2006 Apr  9 & 3835.23557962  &   800     & 33 & SAAO 1.9-m \\
                & 2006 Apr 10 & 3836.29186014  &   800     & 27 & SAAO 1.9-m \\
H$\alpha$103135 & 2007 Mar  7 & 4167.70237491  &   700     & 20 & ESO NTT    \\
                & 2007 Mar  8 & 4168.65039829  &   700     & 14 & ESO NTT    \\
H$\alpha$103959 & 2007 Jan 31 & 4132.53336689  &   800,1000&  9 & SAAO 1.9-m \\
                & 2007 Feb  2 & 4134.42759928  &  1000     & 17 & SAAO 1.9-m \\
                & 2007 Feb  3 & 4135.53816631  &  1000     &  8 & SAAO 1.9-m \\
                & 2007 Feb  4 & 4136.41314177  &  1000     &  5 & SAAO 1.9-m \\
                & 2007 Feb  5 & 4137.40624764  &  1000     & 20 & SAAO 1.9-m \\
H$\alpha$112921 & 2007 Feb  7 & 4139.57558033  &  1000     &  6 & SAAO 1.9-m \\
                & 2007 Feb  8 & 4140.41105671  &  1000     & 19 & SAAO 1.9-m \\
                & 2007 Feb 12 & 4144.51294292  &  1000     &  6 & SAAO 1.9-m \\
                & 2007 Feb 13 & 4145.53923773  &   900,1000&  9 & SAAO 1.9-m \\
                & 2007 Feb 20 & 4152.57276633  &   900     & 11 & CTIO 1.5-m \\
                & 2007 Mar 31 & 4191.56402909  &   900     & 21 & CTIO 1.5-m \\
H$\alpha$122105 & 2007 Mar  4 & 4164.76669129  &   800     & 14 & ESO NTT    \\
                & 2007 Mar  5 & 4165.78076644  &   800     &  7 & ESO NTT    \\
H$\alpha$130559 & 2007 Mar  6 & 4166.72655459  &   700     & 21 & ESO NTT    \\
                & 2007 Mar  8 & 4168.79295985  &   700     & 14 & ESO NTT    \\
H$\alpha$163447 & 2006 Apr  9 & 3835.56287275  &  1000,1200&  8 & SAAO 1.9-m \\
  \hline
  \end{tabular}
\end{table*}

\begin{figure*}
 $\begin{array}{c@{\hspace{2mm}}c}
 \includegraphics[width=88mm]{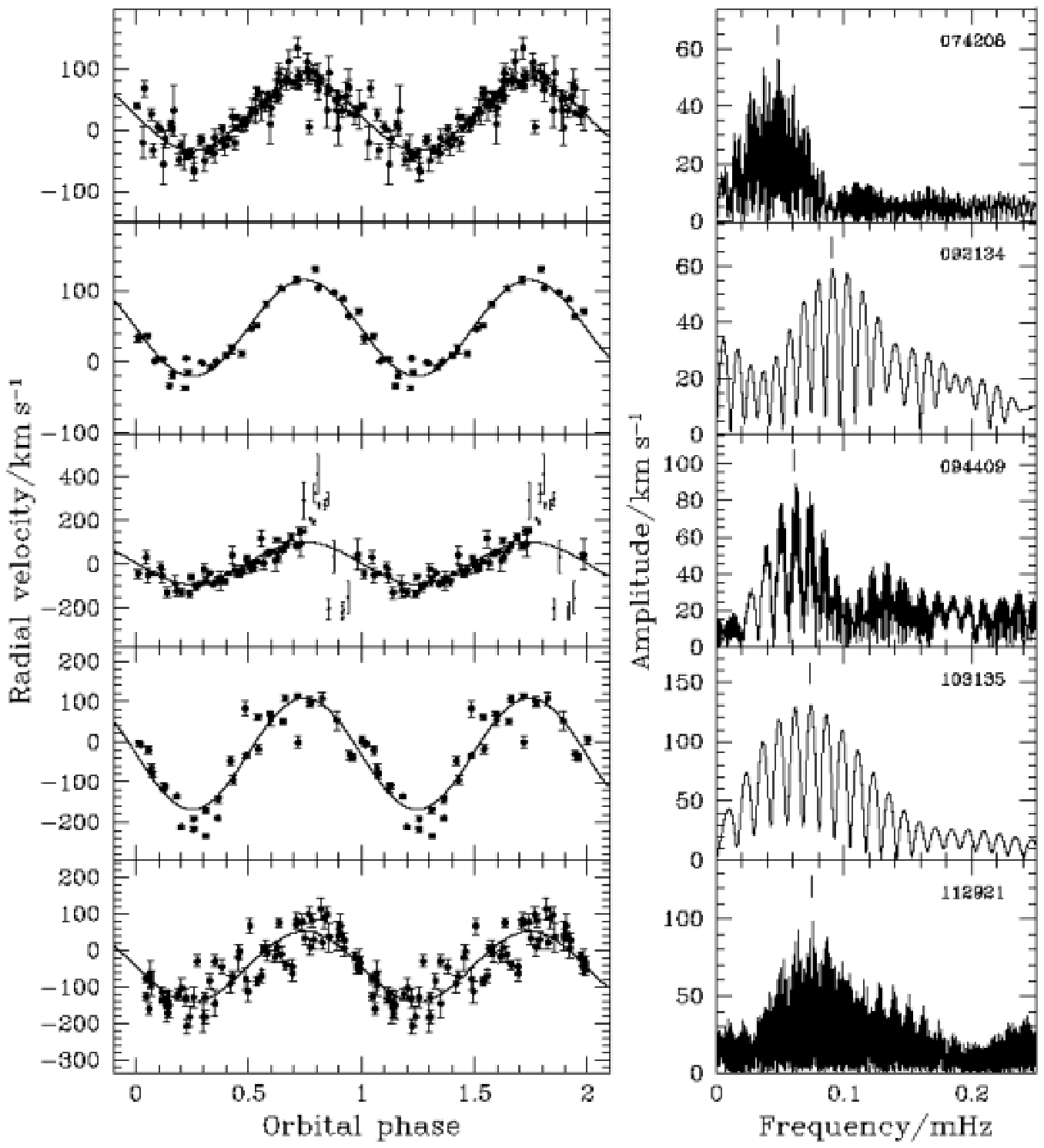} &
 \includegraphics[width=88mm]{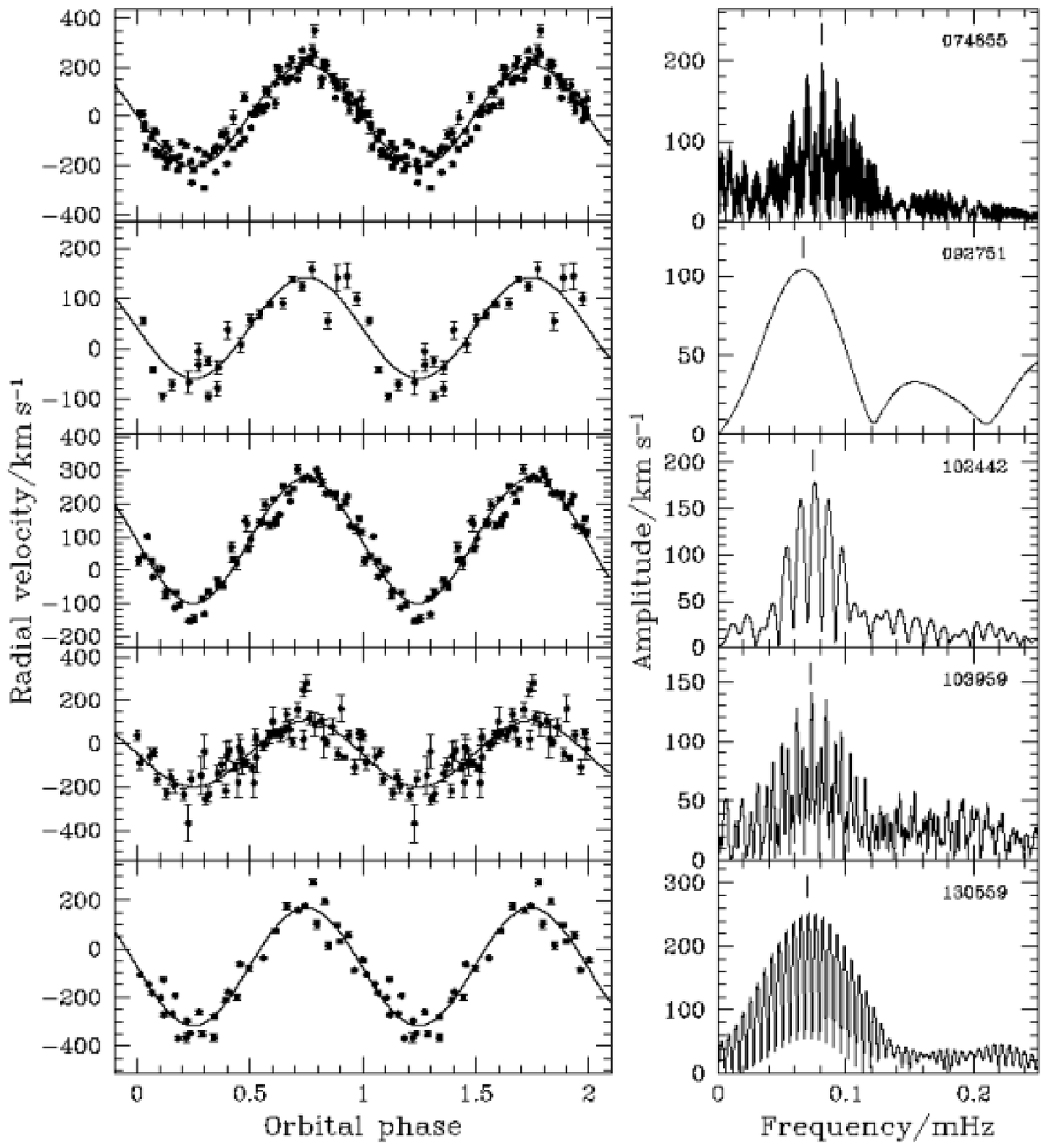} 
 \end{array}$
 \caption {Fourier transforms and phase-folded radial velocity curves.  Vertical bars in the Fourier transforms mark the periods used to fold the data.  One cycle is repeated in the folded radial velocity curves, and the superimposed sine functions are least squares fits to the (unfolded) radial velocities.  
}
 \label{fig:rvandfts}
\end{figure*}

We computed radial velocities from the H$\alpha$ lines, using the Fourier cross correlation method described by \cite{fxcor}, as implemented in the {\sc fxcor} routine in {\sc iraf}.  The radial velocities are found by correlating spectra with a template.  The correlation is restricted to the wavelength range 6\,450 to 6\,650~\AA.  The template spectrum used for the measurements of radial velocities of a given CV is made, over a few iterations, by shifting all individual spectra of that system to 0 velocity (using the velocity measured in the previous iteration), and averaging them.  In the cases where this method did not give satisfactory results (see e.g. Section~\ref{sec:ha074655} below), we used the double Gaussian technique of \cite{SchneiderYoung80}, as implemented in the program {\sc molly}, written by Tom Marsh.  Both methods give formal errors that appear to be too small in several cases (see Fig.~\ref{fig:rvandfts}), but this is not a serious concern, since the amplitudes of the orbital modulations are large compared to the uncertainty indicated by the scatter in radial velocity around these modulations.

We succeeded in measuring spectroscopic orbital periods for 10 systems.  Fig.~\ref{fig:rvandfts} shows Fourier transforms of the radial velocities of these CVs, together with phase-folded radial velocity curves.  The largest amplitude signal in every Fourier transform (marked by vertical bars in Fig.~\ref{fig:rvandfts}) was assumed to represent the orbital modulation of the system, although most of the data sets are actually aliased.  These periods were used to phase-fold the data.

A function of the form 
\begin{equation}
v(t)=\gamma - K \sin \left [ 2\pi \left ( t-T_0/P_{orb} \right ) \right ]
\end{equation}
was fitted to the radial velocity curves by least squares.  $T_0$ is the epoch of red to blue crossing of the radial velocity (i.e. inferior conjunction of the secondary, if the emission lines trace the motion of the white dwarf).  The fits are shown with the data in Fig.~\ref{fig:rvandfts}.  The radial velocities of CV emission lines usually do not trace the dynamical motion of the white dwarf reliably, probably because of contamination from higher velocity components (e.g. \citealt{Smak69}; \citealt{Stover81}; \citealt{MarshHorneShipman87}; \citealt{HessmanKoesterSchoembs89}).  However, since all velocity components are nevertheless modulated at the orbital period, and since this is the only binary parameter we want to measure, we have made no serious attempt to find the true velocity of the primary.  

The radial velocity curve of H$\alpha$094409 has a rotational disturbance near phase 0.  The points that are plotted as smaller symbols in Fig.~\ref{fig:rvandfts} were excluded from both the Fourier transform and the fit.  For all other systems we have used all observations listed in Table~\ref{tab:tr_spect}.

\begin{figure*}
 $\begin{array}{c@{\hspace{2mm}}c@{\hspace{2mm}}c@{\hspace{2mm}}c@{\hspace{2mm}}c}
 \includegraphics[width=32mm]{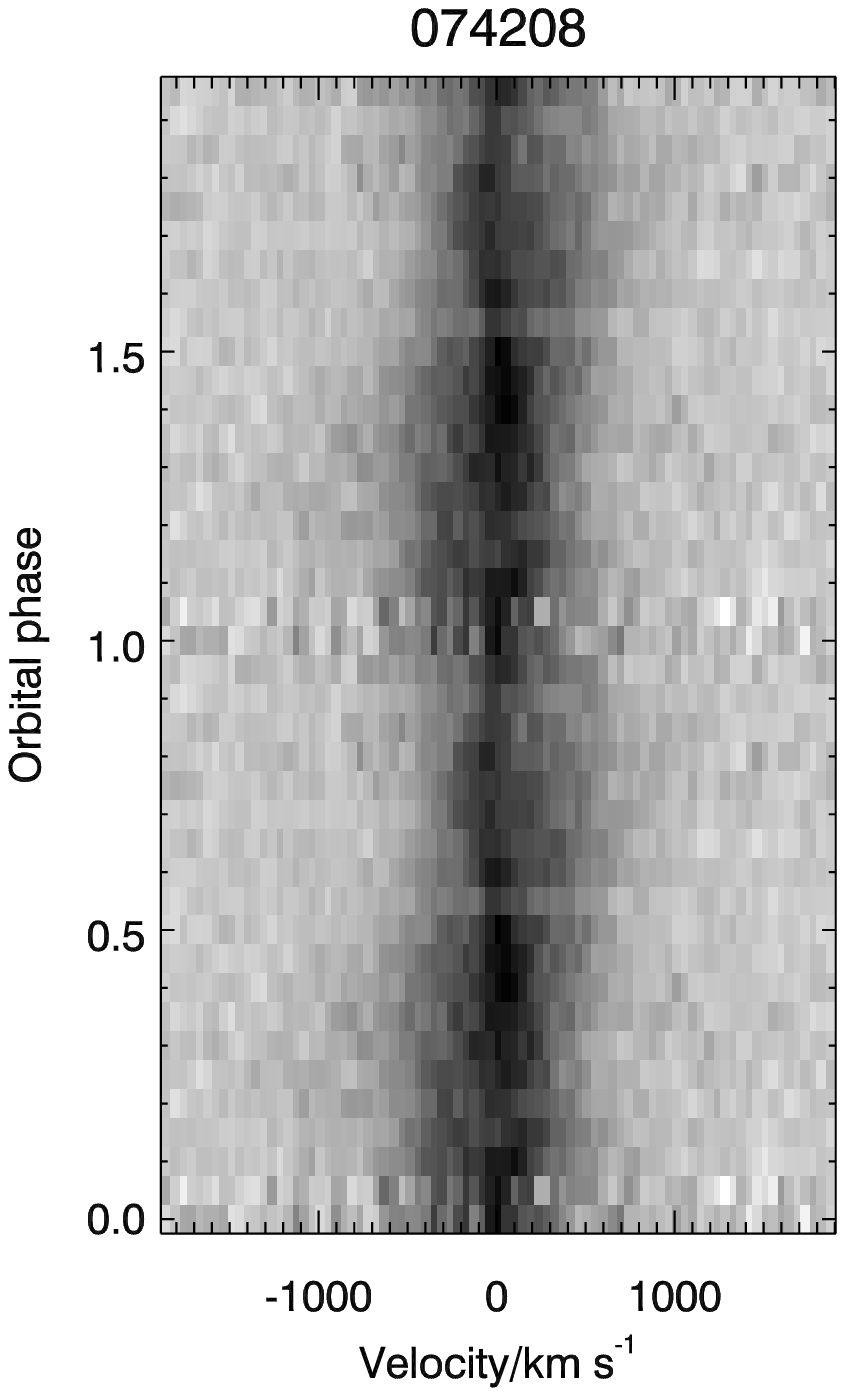} &
 \includegraphics[width=32mm]{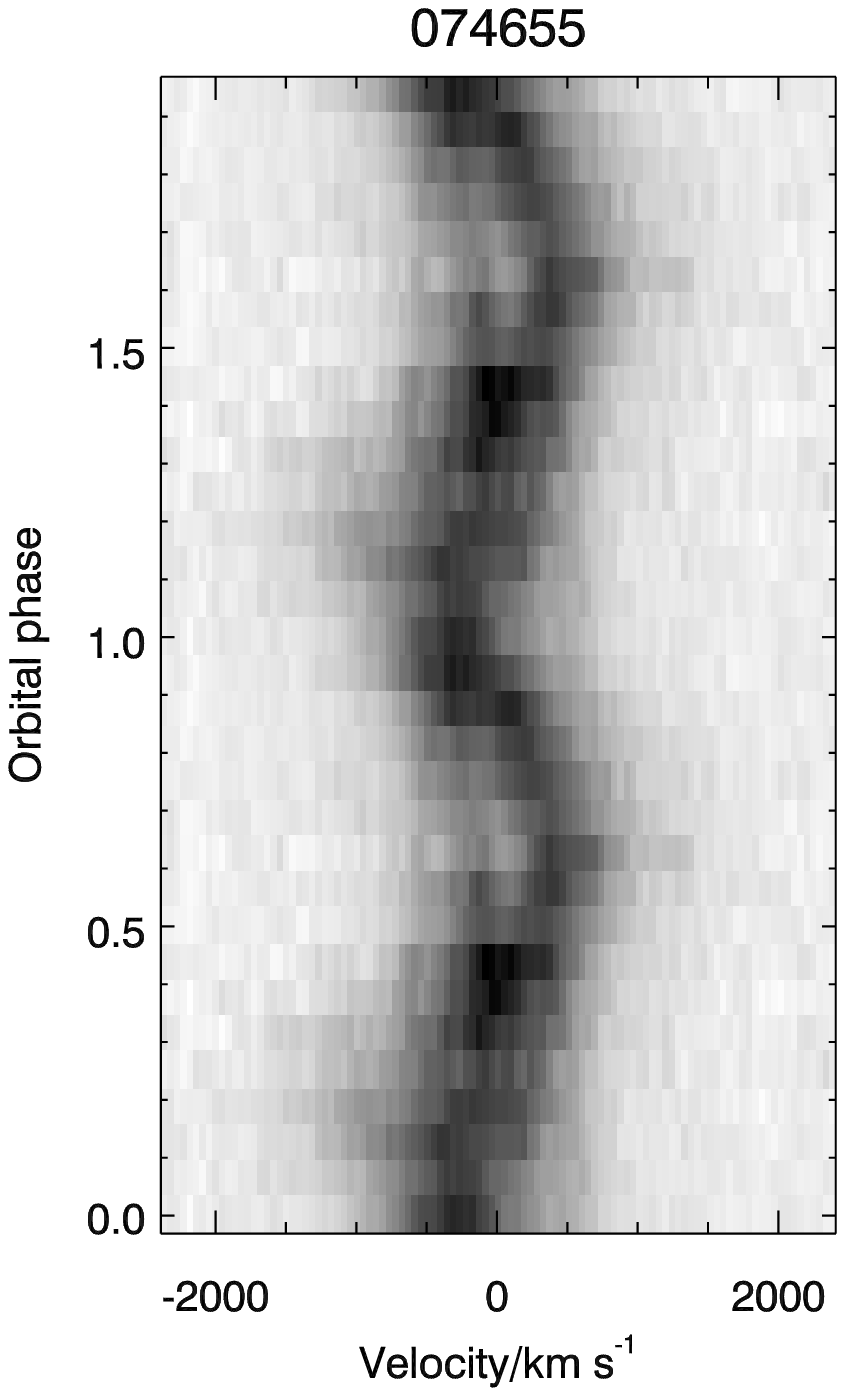} &
 \includegraphics[width=32mm]{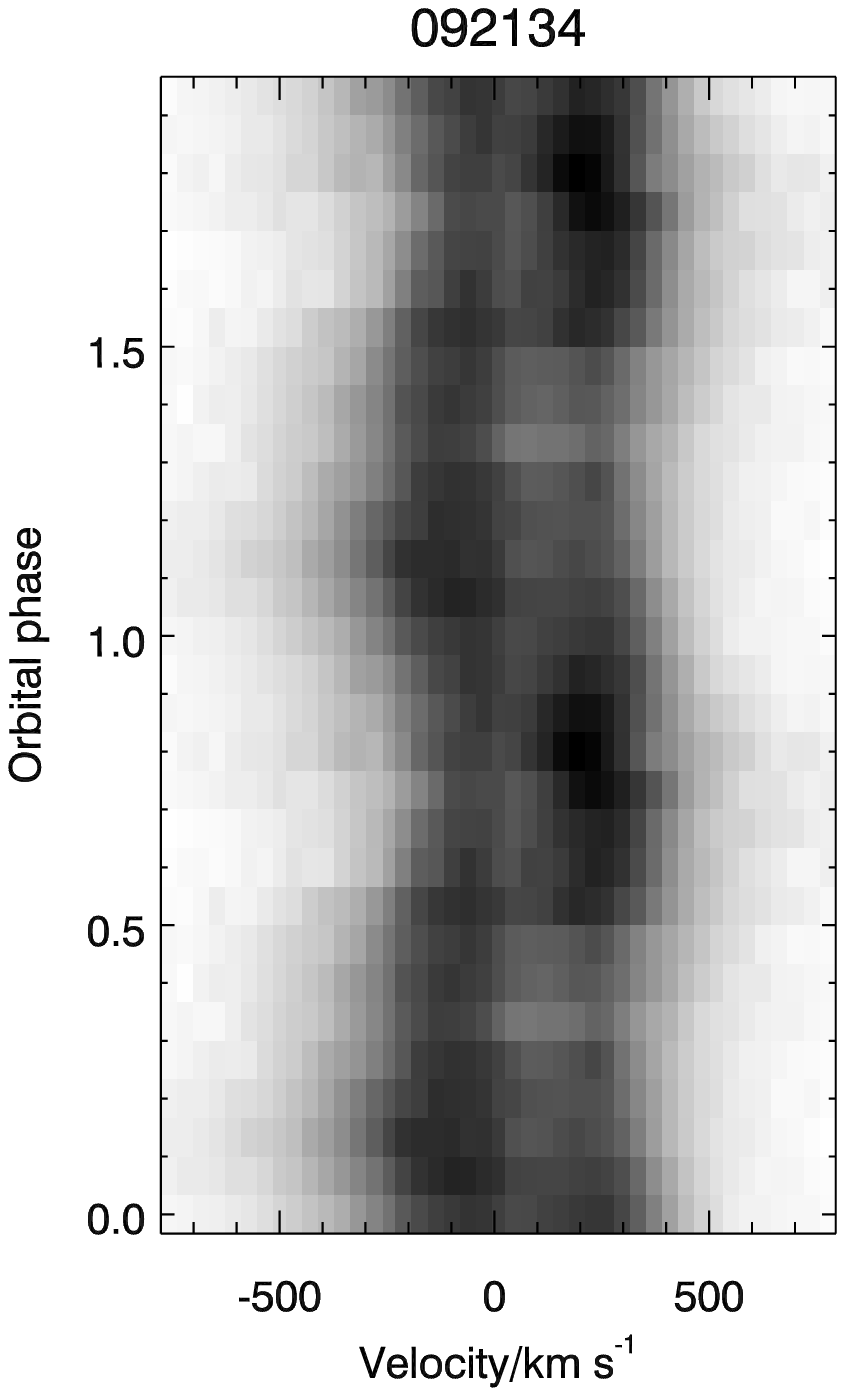} &
 \includegraphics[width=32mm]{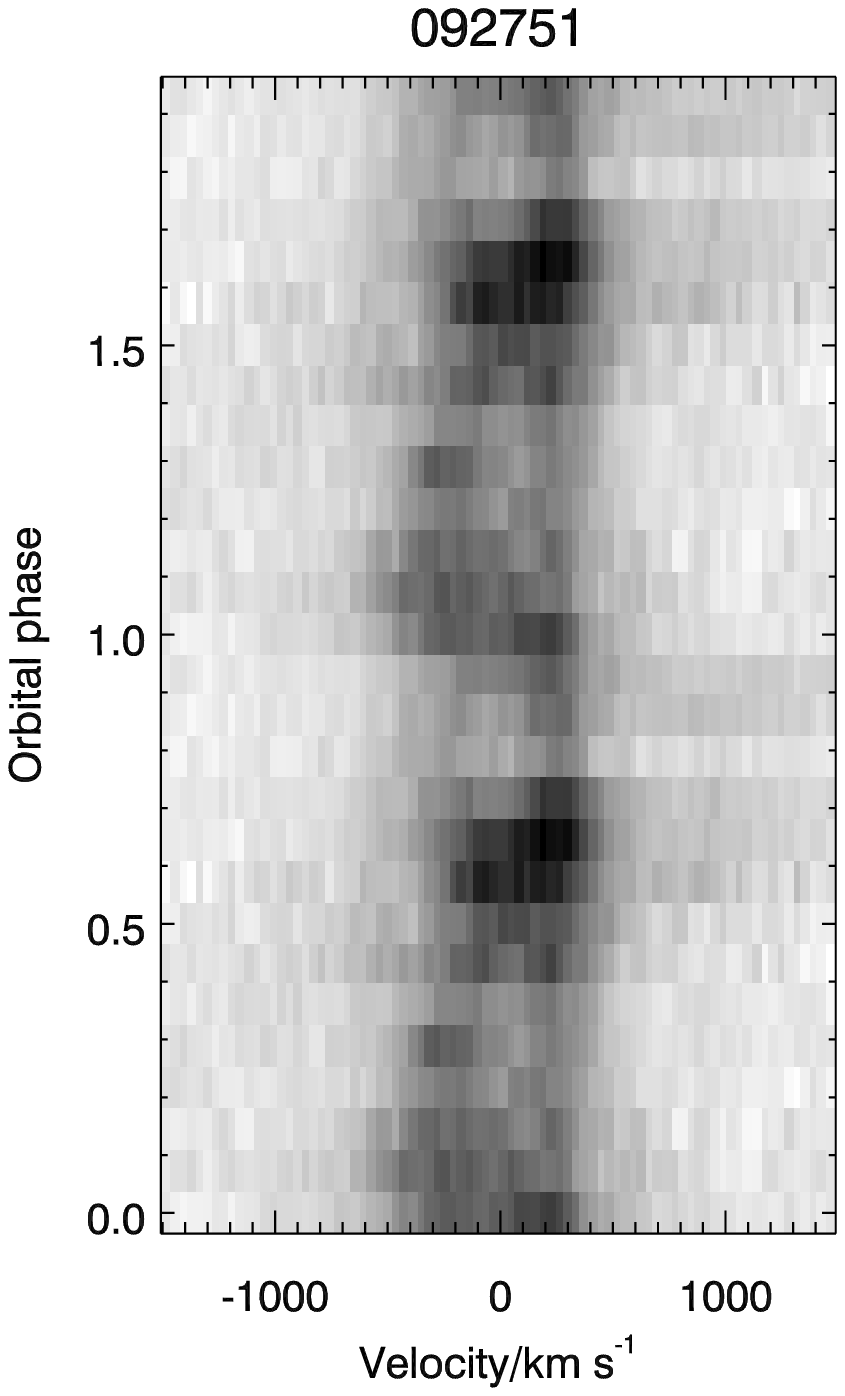} &
 \includegraphics[width=32mm]{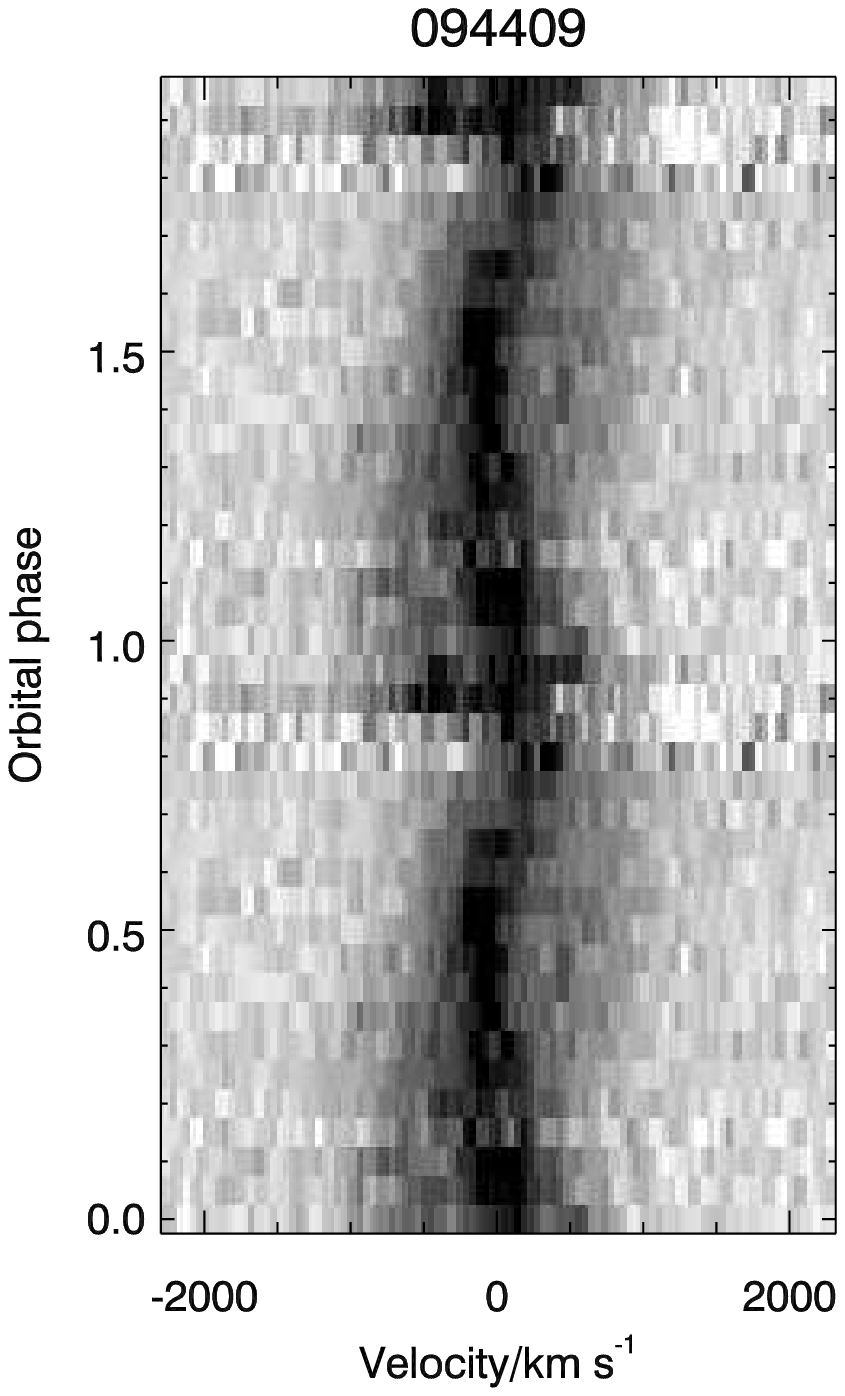} \\
 \includegraphics[width=32mm]{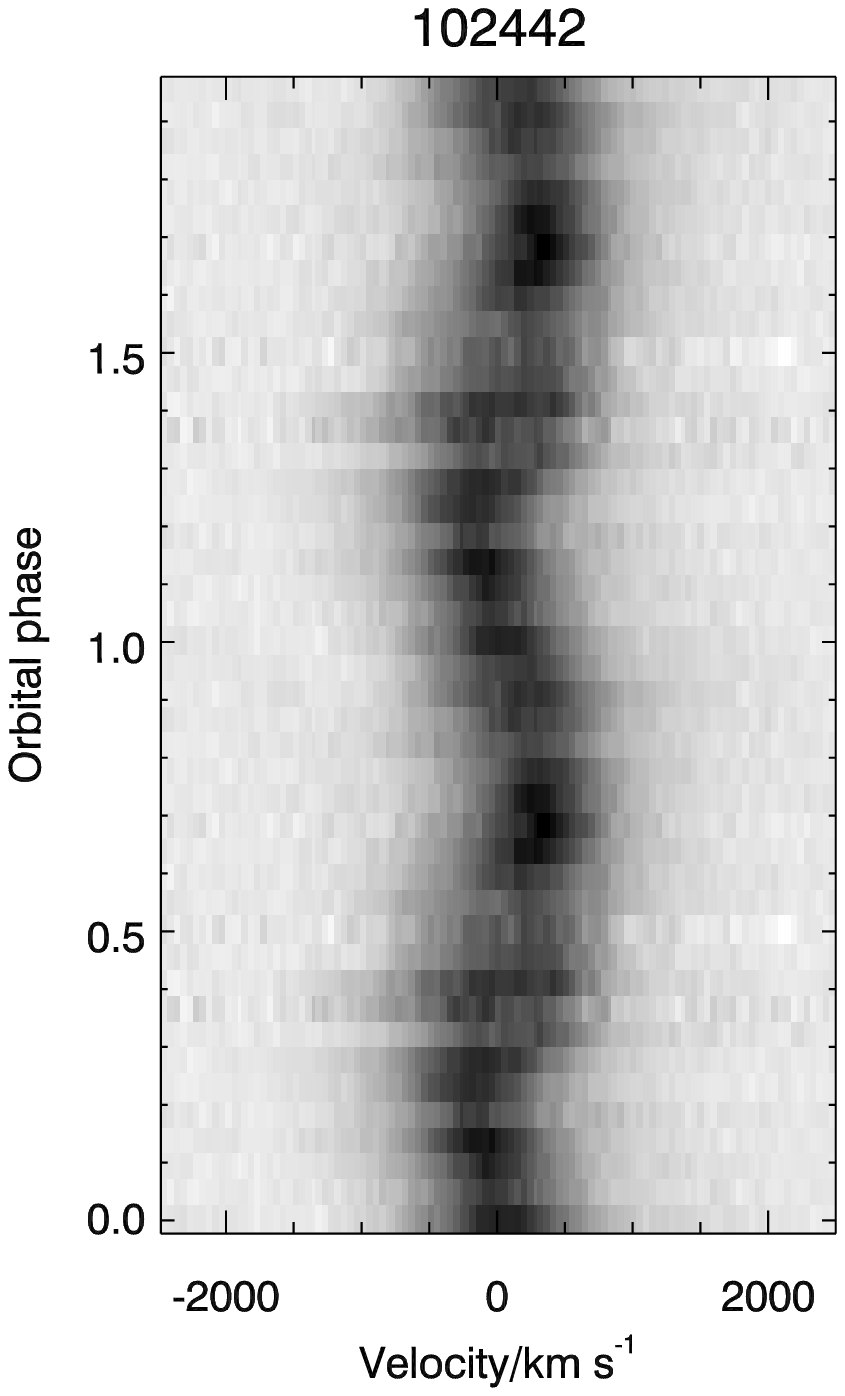} &
 \includegraphics[width=32mm]{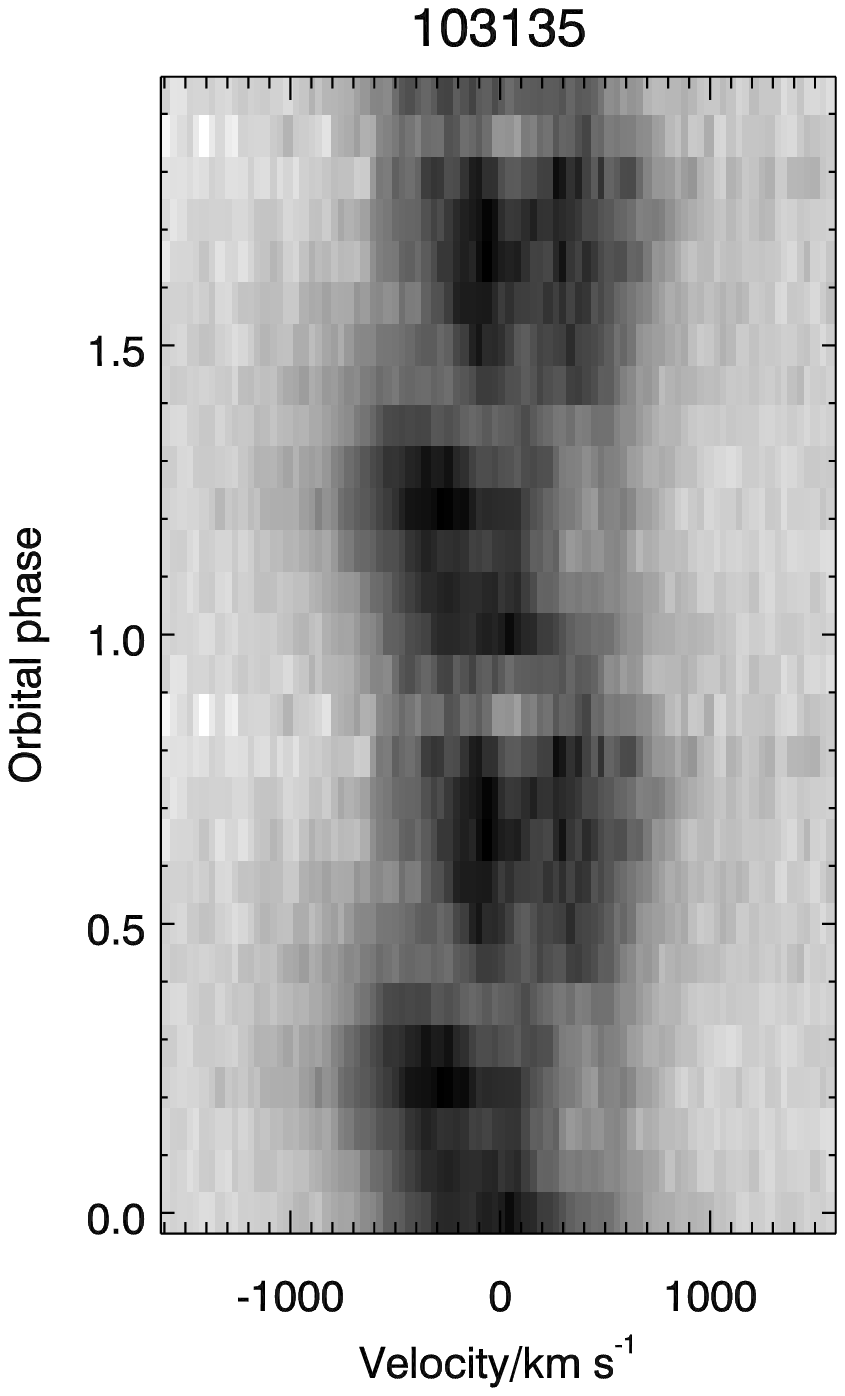} &
 \includegraphics[width=32mm]{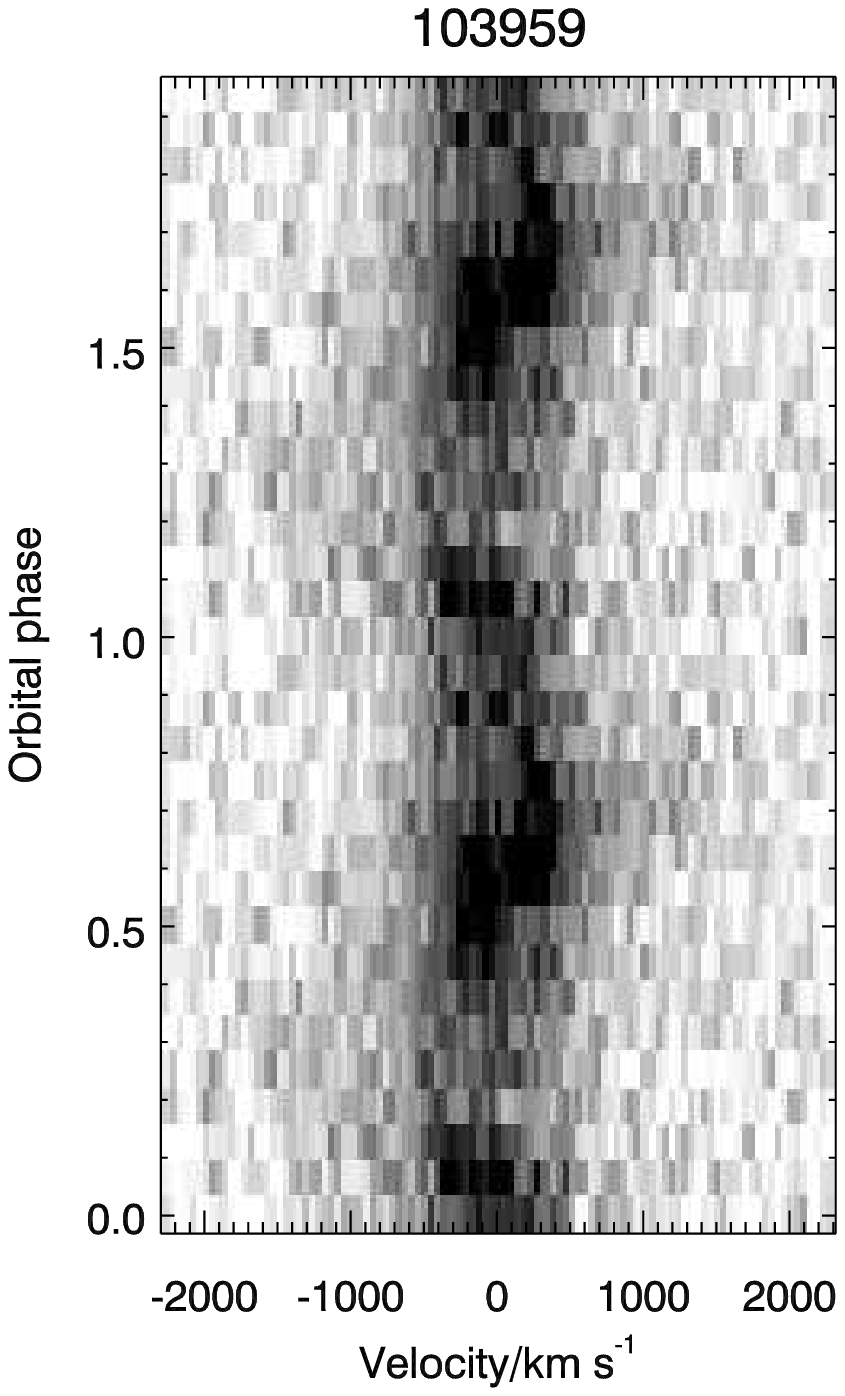} &
 \includegraphics[width=32mm]{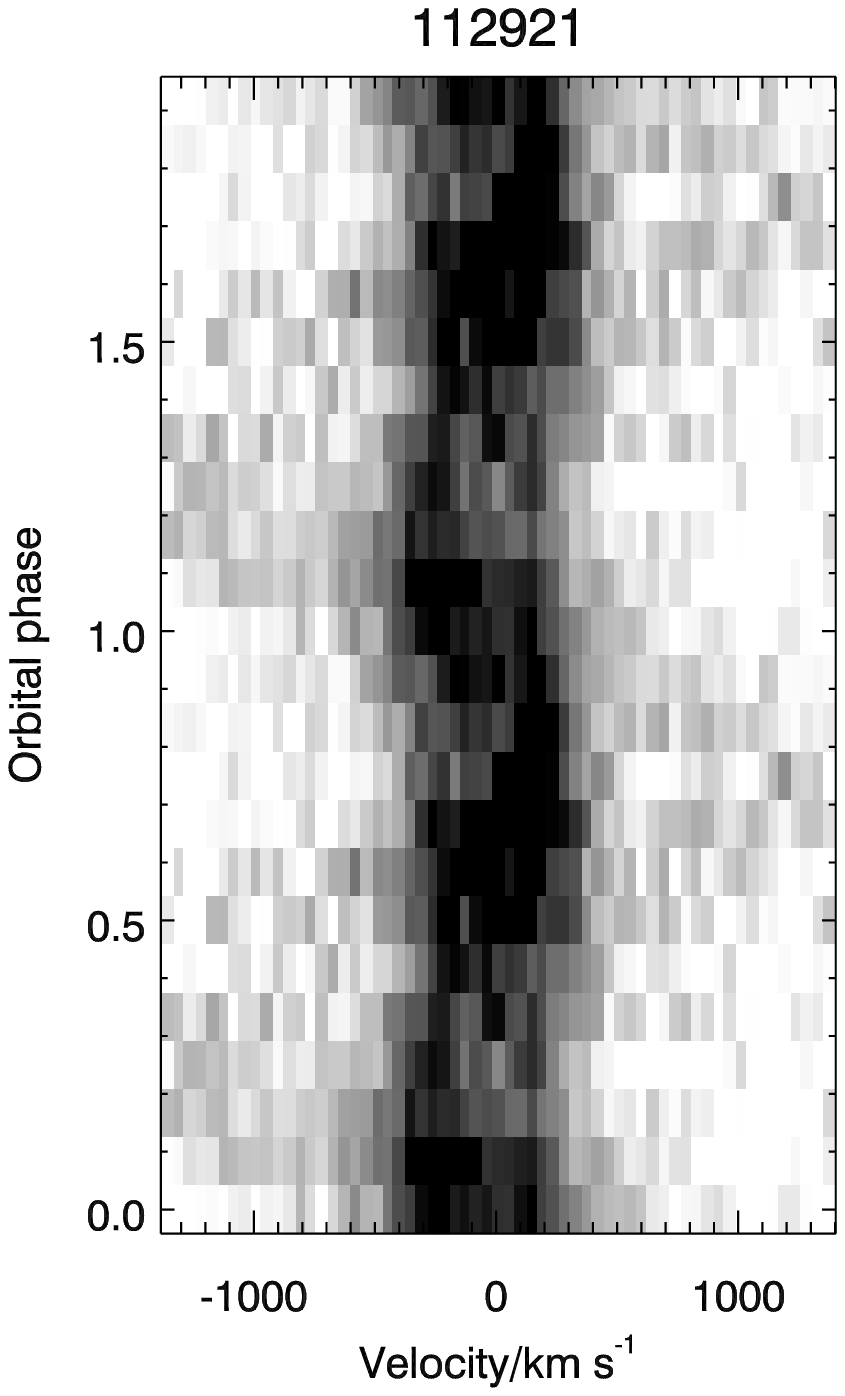} &
 \includegraphics[width=32mm]{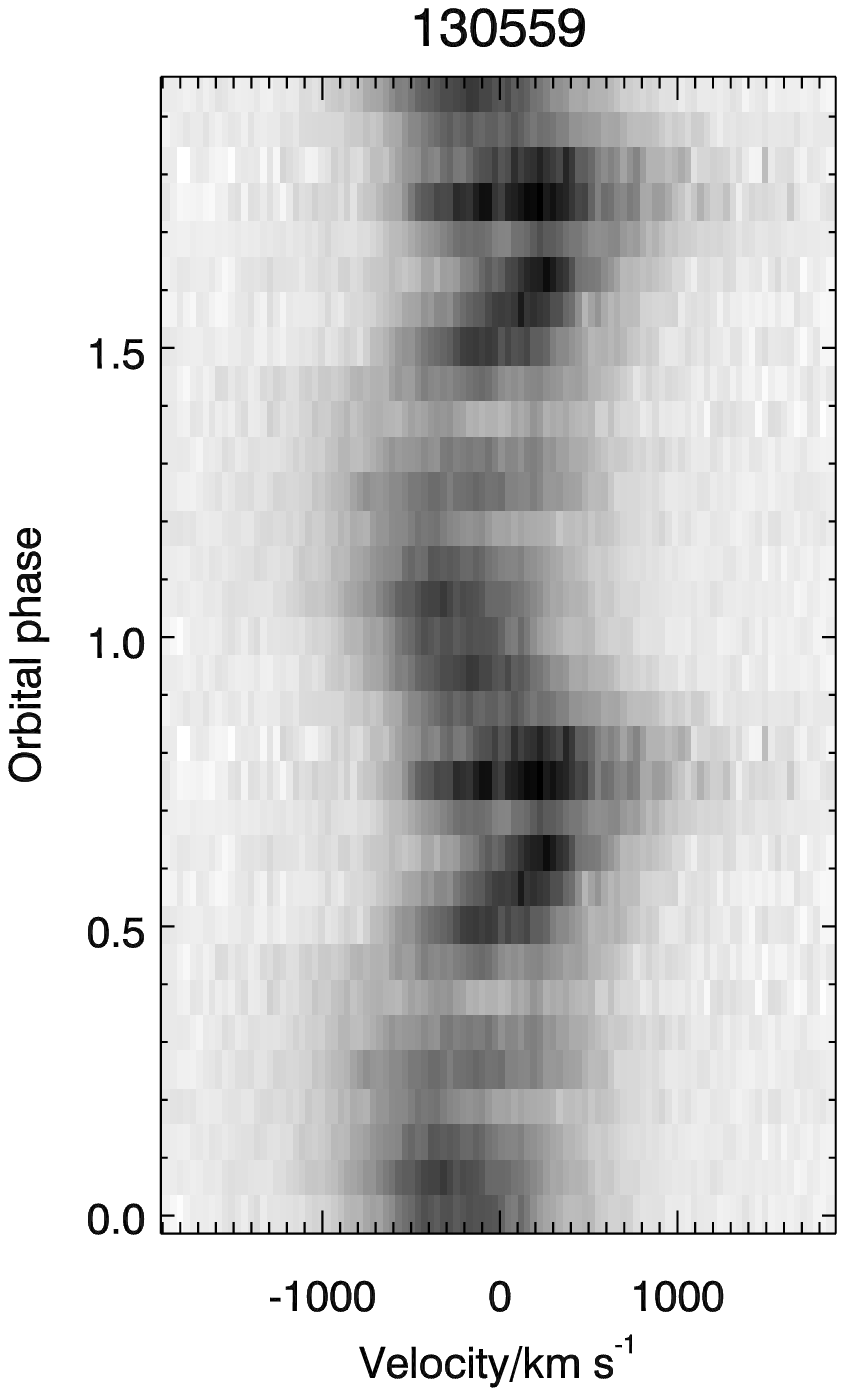}   
 \end{array}$
 \caption {Phase folded and binned spectra.  The individual spectra were normalized to the continuum before being binned.  The grey scale is linear, with higher flux being darker.  Minimum and maximum threshold value were set for some of the images, to enhance the contrast.
}
 \label{fig:trails}
\end{figure*}

Phase binned trailed spectra are shown in Fig.~\ref{fig:trails}.  Velocities are relative to the rest wavelength of H$\alpha$.  For the systems where we have time-resolved spectra from both SAAO and CTIO, data from only one of the telescopes were used to make the trailed spectra (SAAO data for H$\alpha$074208 and H$\alpha$112921, and CTIO data for H$\alpha$074655).  Individual spectra were continuum-normalized before being folded and binned.

\begin{figure*}
 \includegraphics[width=178mm]{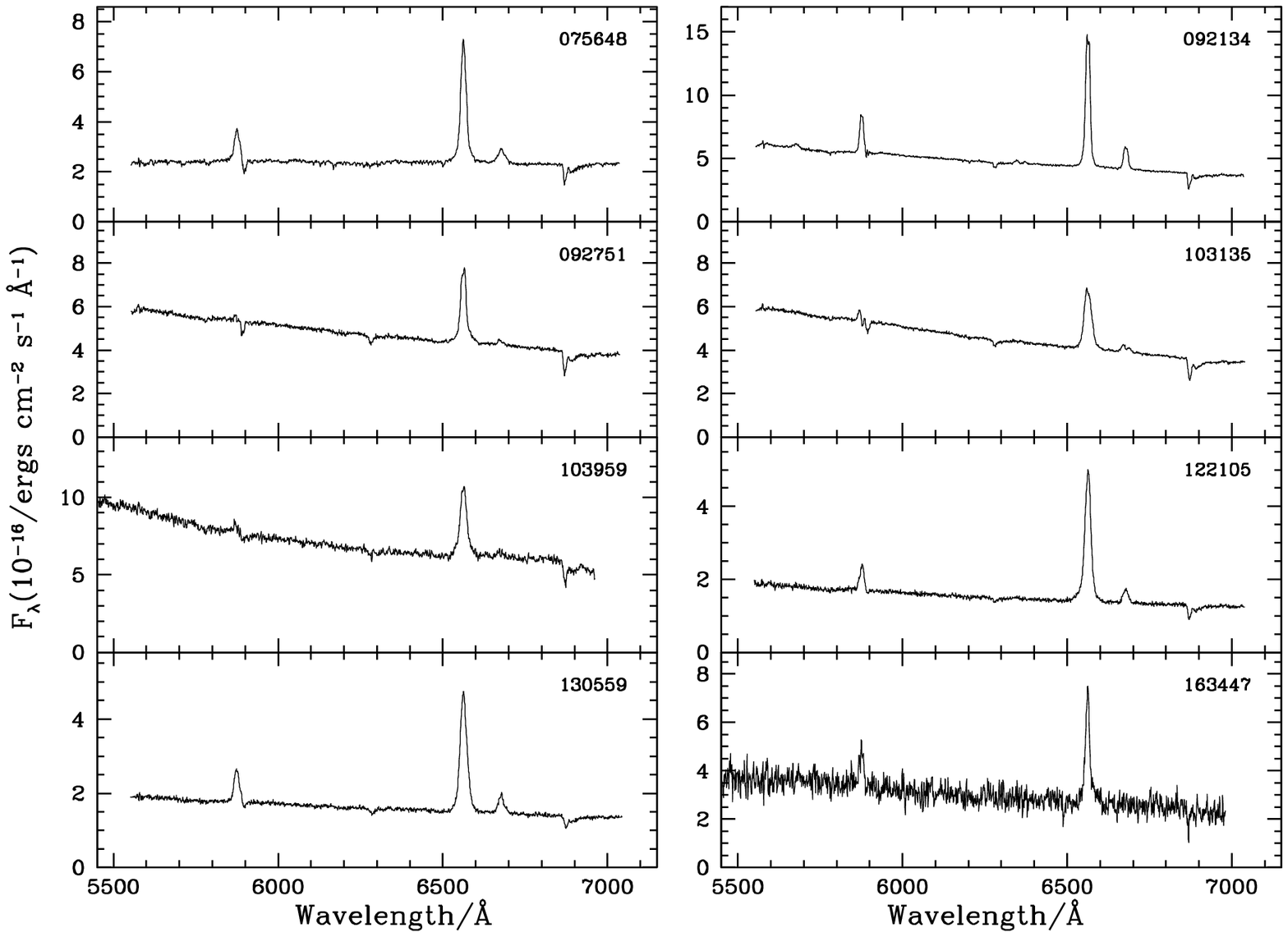}
 \caption{The averages of time-resolved spectra for the systems with the lowest $S/N$ discovery observations, and for which we obtained additional observations.  Individual spectra were shifted to the rest frame using the measured radial velocities before being averaged.
}
 \label{fig:avspectra}
\end{figure*}

Fig~\ref{fig:avspectra} displays average spectra for some of the objects with low $S/N$ identification observations.  These spectra were made by averaging all the time-resolved spectra of a given system, after shifting each spectrum to the rest frame.

\subsection{Time-resolved photometry}
In addition to the time-resolved spectroscopy, we obtained high-speed photometry of several of the new CVs, using the University of Cape Town CCD photometer (UCT CCD; see \citealt{uctccd}) on the SAAO 1-m and 0.76-m telescopes.  Table~\ref{tab:hs_phot} gives a log of the photometry.  These observations were made in white light.  With the UCT CCD, unfiltered observations give photometry with an effective wavelength similar to Johnson $V$, but with a very broad bandpass.  The non-standard flux distribution of CVs and the use of white light means that the observations cannot be precisely placed on a standard photometric system.  Observations of standard stars allow for a magnitude calibration that approximates Johnson $V$ to within $\simeq 0.1$~mag.  We performed differential photometry, implying that colour differences between our targets and the comparison stars were ignored in correcting the photometry for atmospheric extinction.  

\begin{table*}
 \centering
  \caption{Log of the high-speed photometry.  Dates are for the start of the night, and $t_{int}$ is the integration time (the photometer is a frame transfer CCD, so that there is no dead time between exposures).}
  \label{tab:hs_phot}
  \begin{tabular}{@{}llllllll@{}}
  \hline
Object          & Run no. & Date        & HJD $2450000.0 +$ & Run length/h & $t_{int}$/s & telescope & $V$ \\
  \hline
H$\alpha$073418 & RP46    & 2007 Jan 26 & 4127.2845375      & 3.67         & 30         & 1-m     & 18.0$^a$\\
                & RP47    & 2007 Jan 27 & 4128.2790181      & 6.65         & 30         & 1-m     & 17.9$^a$\\
H$\alpha$074655 & RP53    & 2007 Jan 30 & 4131.2769557      & 2.07         &  8         & 1-m     & 14.0 \\
H$\alpha$075648 & RP49    & 2007 Jan 28 & 4129.2811127      & 1.90         & 10         & 1-m     & 16.6 \\
H$\alpha$092134 & RP50    & 2007 Jan 28 & 4129.3580250      & 2.11         & 10         & 1-m     & 17.0 \\
H$\alpha$092751 & RP51    & 2007 Jan 28 & 4129.4516001      & 1.85         & 10         & 1-m     & 17.8 \\
H$\alpha$094409 & RP54    & 2007 Jan 30 & 4131.3676657      & 1.52         & 10         & 1-m     & 16.7:$^a$\\
H$\alpha$102442 & RP41    & 2006 Apr  8 & 3834.3277709      & 3.58         & 10         & 0.76-m  & 16.2 \\
                & RP42    & 2006 Apr  9 & 3835.2742616      & 5.33         & 10         & 0.76-m  & 16.0 \\
                & RP44    & 2006 Apr 10 & 3836.2993854      & 3.40         & 10         & 0.76-m  & 16.0 \\
H$\alpha$103135 & RP52    & 2007 Jan 28 & 4129.5336457      & 2.30         & 10         & 1-m     & 17.2:\\
H$\alpha$103959 & RP45    & 2007 Jan 24 & 4125.5677782      & 1.18         & 10         & 1-m     & 16.3:\\
                & RP48    & 2007 Jan 27 & 4128.5618716      & 1.67         & 10         & 1-m     & 16.2 \\
H$\alpha$122105 & RP31    & 2005 Jun 10 & 3532.2414425      & 4.42         & 30         & 1-m     & 17.5 \\
                & RP34    & 2005 Jun 11 & 3533.2653467      & 2.48         & 20,30      & 1-m     & 17.5 \\
H$\alpha$163447 & RP43    & 2006 Apr  9 & 3835.5055942      & 4.12         & 10,14      & 0.76-m  & 18.3 \\
  \hline
  \end{tabular}
\\
Notes: `:' denotes an uncertain value; $^a$average magnitude outside of eclipse. \hfill
\end{table*}

\begin{figure*}
 \includegraphics[width=178mm]{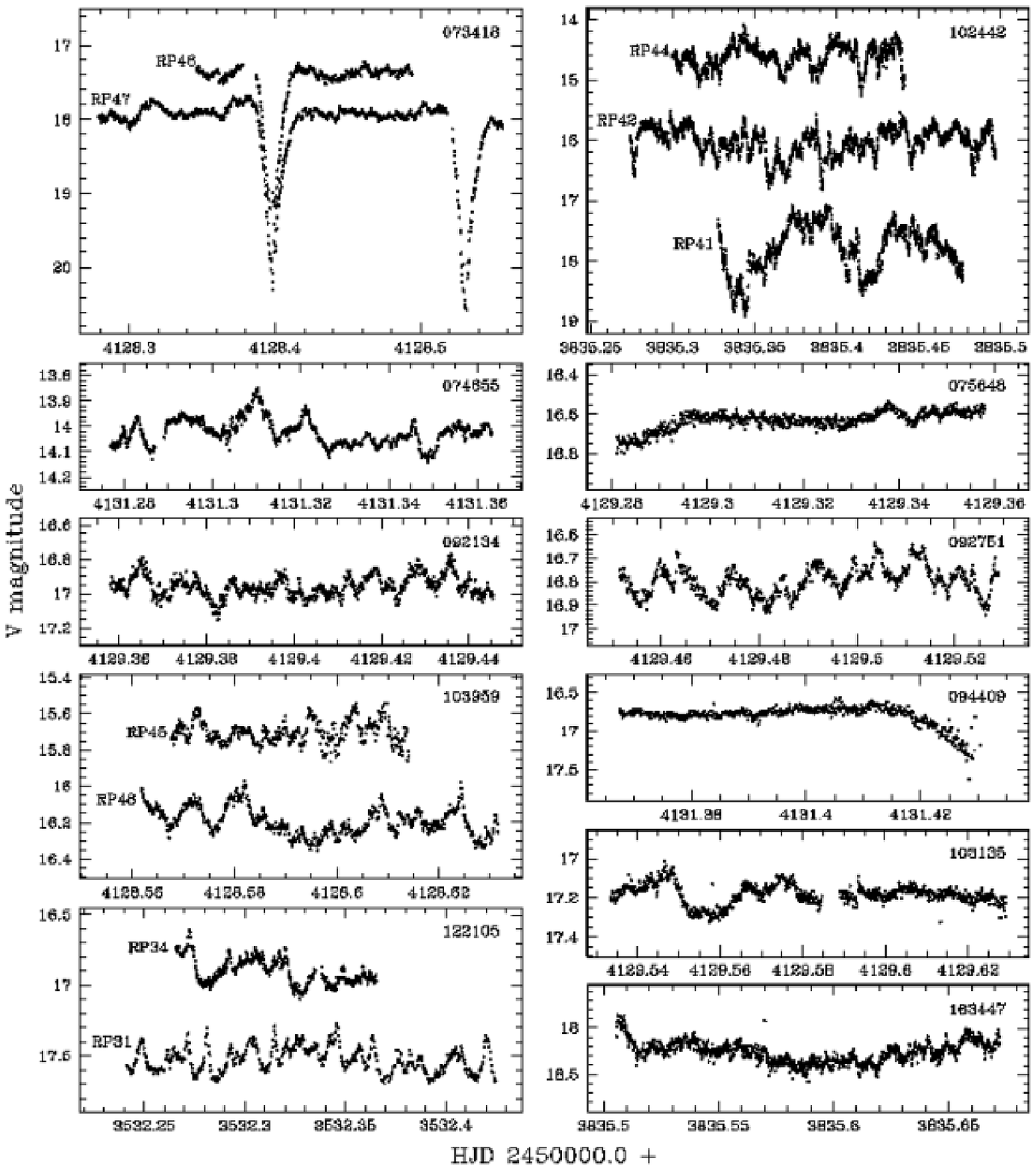}
 \caption{Light curves for 11 of the new CVs.  Details of the observations are given in Table~\ref{tab:hs_phot}.  For the systems with observations on more than one night (H$\alpha$073448, H$\alpha$102442, H$\alpha$103959, and H$\alpha$122105), arbitrary shifts were applied to one or two of the light curves, for display purposes, as specified in the text.
}
 \label{fig:lightcurves}
\end{figure*}

The light curves are shown in Fig.~\ref{fig:lightcurves}.  The objects all display rapid flickering (the observational signature of mass transfer), and both H$\alpha$073418 and H$\alpha$094409 are found to be eclipsing systems.  All light curves were searched for rapid oscillations, in part because we expected that a few of the objects might be intermediate polars.  We found no compelling evidence for coherent modulations on short time-scales in any of the photometry, but there is some indication of a $\simeq 15\,\mathrm{min}$ QPO (e.g. \citealt{PattersonRobinsonNather77}; \citealt{Warner04}) in one of the light curves of H$\alpha$122105 (see Section~\ref{sec:075648and122105}).

\section{Results for individual systems}
\label{sec:results}

\subsection{H$\alpha$073418}
The identification spectrum of H$\alpha$073418 shows single-peaked Balmer, as well as He\,{\scriptsize I} and He\,{\scriptsize II}\,$\lambda$4686, emission lines on a reasonably flat continuum.  

We obtained high-speed photometry on two consecutive nights, and found that H$\alpha$073418 is a deeply eclipsing system.  The light curves are displayed in Fig.~\ref{fig:lightcurves}; the run RP46 data are shifted vertically by $-0.6$, and horizontally by 8 orbital cycles.  We were able to determine the cycle count over the two nights unambiguously using the eclipse timings.  To derive an ephemeris, we fitted a line to the times of minimum by least squares.  The ephemeris of mid-eclipse is 
\begin{equation}
\mathrm{HJD_{min}}=2\,454\,127.33644(2)+0.132726(2)\mathrm{E}.
\label{eq:eph}
\end{equation}

The absence of an orbital hump implies a high-$\dot{M}$ disc, and the single-peaked spectral lines and V-shaped eclipse profile suggest that H$\alpha$073418 might be an SW Sex star (e.g. \citealt{ThorstensenRingwaldWade91}; \citealt{RuttenvanParadijsTinbergen92}; \citealt{Horne99}; \citealt{KniggeLongHoard00}).  The orbital period also fits in with this classification, since SW Sex stars are concentrated in the narrow period range 3--4~h (\citealt{ThorstensenRingwaldWade91} already noted that SW Sex stars occupy a narrow range in $P_{orb}$; \citealt{Rodriguez-Gil07} recently argued that they are the dominant CV population in this period range).  However, without time-resolved spectroscopy it is not possible to classify a CV as an SW Sex star with any certainty.

\subsection{H$\alpha$074208}
H$\alpha$074208 is probably the optical counterpart of the \emph{ROSAT} source 1RXS J074206.4-104929.  It has H\,{\scriptsize I}, He\,{\scriptsize I}, and He\,{\scriptsize II} emission lines, and a continuum that is red at wavelengths above $\simeq5\,000~\,\mathrm{\AA}$, and blue below $\simeq4\,200~\,\mathrm{\AA}$.  Absorption bands from a late-type secondary star are also clearly detected.

\begin{figure}
 \includegraphics[width=84mm]{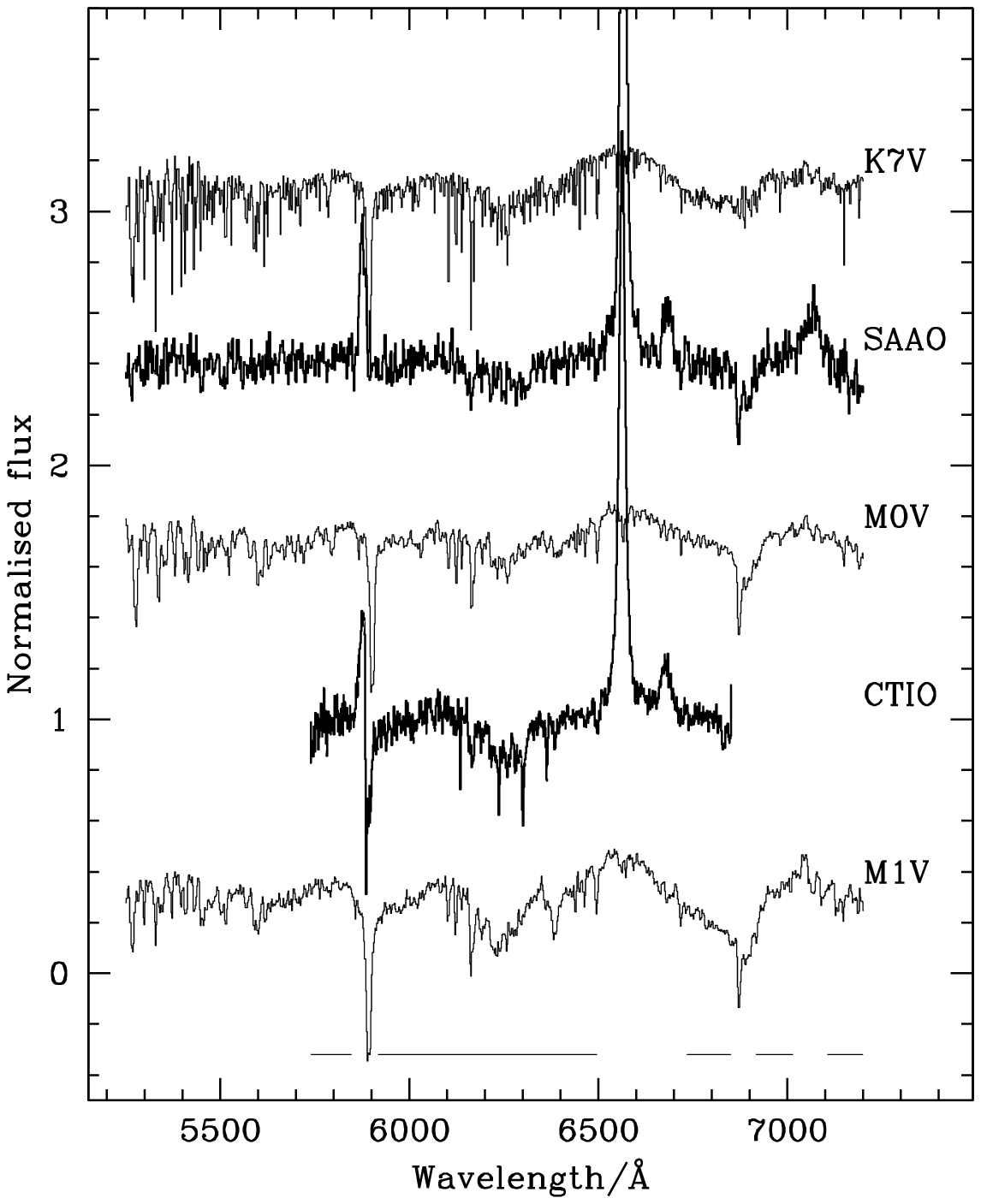}
 \caption{A comparison of the observed spectra of H$\alpha$074208 with late-type dwarf templates.  All the spectra are normalized, and arbitrarily offset.  The bold spectra are the SAAO discovery spectrum and the average of all spectra obtained at CTIO.
}
 \label{fig:Ha074208_secondary}
\end{figure}

We did not obtain any high-speed photometry of H$\alpha$074208.  Time-resolved spectra were taken at both CTIO and SAAO.  The Fourier transform of the radial velocity curve of the H$\alpha$ line is shown, together with the phase folded radial velocities, in Fig.~\ref{fig:rvandfts}.  From these data we measure $P_{orb}=5.706\pm0.003\,\mathrm{h}$; the cycle count is unlikely to be wrong.  We were not able to measure the radial velocity of the secondary star.

The red end of the spectrum can be used to measure the spectral type of the secondary.  We find a rough estimate of M$0\pm1$, by visually comparing the spectra to templates from \cite{JacobyHunterChristian84} and \cite{Valdes04}\footnote{Note that the spectral classes M1 and M0 appear to have been inverted in \cite{JacobyHunterChristian84}.  The spectrum classified as M0V is clearly later than the one labelled M1V in their fig. 2g.}.  Fig.~\ref{fig:Ha074208_secondary} shows the observed spectra, together with three templates.  

We make the (very simplistic) assumption that the fractional flux contribution of other components is constant with wavelength over the continuum bands indicated by horizontal bars along the bottom of Fig.~\ref{fig:Ha074208_secondary}, and find that the secondary contributes between roughly 40 and 80\% of the total flux at those wavelengths.  The larger secondary contributions are found using the CTIO data.  Our spectra are not well suited to measuring broad-band magnitudes, since the wavelength coverage is not wide enough, and the absolute flux calibration is not very reliable.  However, we can estimate rough $R$-band magnitudes of $15.8\pm0.2$ and $16.1\pm0.3$ from the SAAO and CTIO data, respectively.  Taking this at face value, the system was fainter when it was observed at CTIO, so that it is understandable that the secondary makes a larger relative flux contribution in those data.  \cite{Knigge06} predicts $M_R=7.3\pm0.4$ for a typical unevolved CV secondary with spectral type between K7 and M1.  Using the extinction models of \cite{DrimmelCabrera-LaversLopez-Corredoira03} and \cite{AmoresLepine05} we estimate $A_R=0.3\pm0.1$, for distances up to between 500 and 1\,000~pc along this line of sight.  This leads to a distance estimate of $580^{+160}_{-110}\,\mathrm{pc}$ for H$\alpha$074208.  There is no direct evidence that the system has an evolved secondary (unevolved CVs at 5.7~h have secondary spectral types of about M0.7; \citealt{Knigge06}).  However, CVs with evolved secondaries probably dominate the population at periods above $\simeq5$~h (\citealt{BeuermannBaraffeKolb98}; \citealt{BaraffeKolb00}; \citealt{PodsiadlowskiHanRappaport03}).  If H$\alpha$074208 is an evolved system, the distance estimate is invalid, since it assumes the luminosity predicted for an unevolved secondary.

\subsection{H$\alpha$074655}
\label{sec:ha074655}
At $V=14.0$, H$\alpha$074655 is the brightest CV in our sample.  The identification spectrum shows the double-peaked emission lines indicative of a high-inclination accretion disc, and a very blue continuum.  In addition to H\,{\scriptsize I} and He\,{\scriptsize I}, we also detect He\,{\scriptsize II}\,$\lambda$4686 and the C\,{\scriptsize III}/N\,{\scriptsize III}\,$\lambda\lambda$4640--4650 Bowen blend in emission.

Probably because of variation in line profile shape as a function of orbital phase (see Fig~\ref{fig:trails}), a cross correlation with a single template spectrum produces a non-sinusoidal radial velocity curve (there is no reason to suspect that the orbit is eccentric).  We therefore measured radial velocities of this system with the double Gaussian technique.  We used a Gaussian FWHM of $300\,\mathrm{km/s}$, and separation of $1\,600\,\mathrm{km/s}$ to obtain the radial velocity curve shown in Fig.~\ref{fig:rvandfts}.  The best fit sinusoid gives a rather high radial velocity amplitude of $208\,\mathrm{km/s}$, which supports the suggestion that it is a high inclination system.  For this object, we are able to discard other aliases as very unlikely to be the correct frequencies.  The orbital period of H$\alpha$074655 is $3.3984\pm0.0004\,\mathrm{h}$.  Our only light curve is shorter than the orbital period, and does not show any obvious orbital modulation.

\subsection{H$\alpha$092134}
The identification spectrum of H$\alpha$092134 has Balmer and He\,{\scriptsize I} emission lines superimposed on a blue continuum.  The data also show strong He\,{\scriptsize II}\,$\lambda$4686 emission.  A higher quality spectrum is displayed in Fig.~\ref{fig:avspectra}.  The light curve of H$\alpha$092134 shows no modulation other than flickering.

Time-resolved spectroscopic observations were taken on two consecutive nights.  The Fourier transform and phase folded radial velocity curve are shown in Fig.~\ref{fig:rvandfts}; note the small radial velocity amplitude (the best fit gives $68.0\,\mathrm{km/s}$).  The strongest power in the Fourier transform of the radial velocity curve is at $9.134 \times 10^{-2}\,\mathrm{mHz}$, corresponding to 3.041~h.  There is some possibility that the period is the next strongest alias (3.479~h).  None of the other aliases provides acceptable fits to the data.  Assuming that the largest amplitude signal represents the orbital modulation, the orbital period is $3.041\pm0.009\,\mathrm{h}$.  

\subsection{H$\alpha$092751}
The spectrum of H$\alpha$092751 has quite weak emission lines superimposed on a blue continuum ($\mathrm{EW(H}\alpha)$ is only $\sim20\,\mathrm{\AA}$, and He\,{\scriptsize I} is barely detected; see Fig.~\ref{fig:idspectra} and \ref{fig:avspectra}).  We obtained one short light curve, which shows the usual short term photometric behaviour of a CV, and is plotted in Fig.~\ref{fig:lightcurves}.

Since we have time-resolved spectroscopy from only one night, we can measure only an imprecise period of $4.1\pm0.3\,\mathrm{h}$.  Our data fortunately sample all orbital phases reasonably well (see Fig.~\ref{fig:rvandfts}).  The amplitude of the best fit to the radial velocities is $100.9\,\mathrm{km/s}$.  The trailed spectrum of H$\alpha$092751, displayed in Fig~\ref{fig:trails}, shows the double-peaked H$\alpha$ profile typical of an accretion disc, and possibly also an S-wave component.

\subsection{H$\alpha$094409}
Balmer, He\,{\scriptsize I}, and strong He\,{\scriptsize II}\,$\lambda$4686 emission lines are detected in the spectrum of H$\alpha$094409.  The lines are single-peaked in the identification spectrum, and the continuum is flat.  Although our only light curve of H$\alpha$094409 was obtained under poor conditions, it shows that this is an eclipsing system.  The eclipses are also seen in the time-resolved spectroscopy, both from the spectrophotometry and from the large rotational disturbance in the radial velocity curve (see Fig.~\ref{fig:rvandfts}).  

If the largest amplitude signal in the Fourier transform of the radial velocity curve corresponds to the orbital modulation, then the orbital period of H$\alpha$094409 is $4.506\pm0.004\,\mathrm{h}$.  The orbital frequency might, however, be identified with several other aliases (the shortest acceptable period is still longer than 3~h).

With phase 0 photometrically determined, the red to blue crossing of the radial velocities is at about phase 0.15.  One of the defining characteristics of SW Sex stars is a phase shift between Balmer emission line velocities and the expected velocity of the white dwarf \citep{ThorstensenRingwaldWade91}.  Two other factors, taken at face value, suggest that H$\alpha$094409 might be an SW Sex star.  The first is that lines are single-peaked in the average spectrum, despite the high binary inclination.  Secondly, the H$\alpha$ line profiles seem to change as a function of orbital phase, with possibly central absorption at some phases (another phenomenon observed in SW Sex stars).  However, the phase shift in the radial velocity curve of H$\alpha$094409 is only marginally significant, since no concerted attempt was made to trace the motion of the white dwarf, and since our low quality eclipse photometry does not allow us to determine phase 0 very accurately.  Thus, while there are some hints that H$\alpha$094409 might be an SW Sex star, our data are not good enough to allow for a firm classification.

\subsection{H$\alpha$102442}
The member of our sample with the strongest Balmer emission is H$\alpha$102442, which has $\mathrm{EW(H}\alpha)\simeq120\,\mathrm{\AA}$.  He\,{\scriptsize I} and He\,{\scriptsize II}\,$\lambda$4686 are also detected in emission.  Time resolved spectra were taken over two nights, and the sampling and data quality were sufficient to establish an unambiguous cycle count between the two radial velocity curves (see the Fourier transform in Fig.~\ref{fig:rvandfts}).  The orbital period of H$\alpha$102442 is $3.673\pm0.006\,\mathrm{h}$.  We measure a quite large radial velocity amplitude (the best-fit amplitude is $188.3\,\mathrm{km/s}$).  The trailed spectrum given in Fig.~\ref{fig:trails} shows a pure S-wave modulation.

We obtained three light curves of this system on consecutive nights, the last two simultaneously with the time-resolved spectroscopy.  These observations are displayed in Fig.~\ref{fig:lightcurves}.  The run RP41 and RP44 data are shifted vertically by $+1.6$ and $-1.4$, respectively, and horizontally by $+1$~d and $-1$~d, respectively.  The photometry shows large amplitude flickering, but nothing that can be recognized as an orbital modulation.

\subsection{H$\alpha$103135}
The identification spectrum of H$\alpha$103135 plotted in Fig.~\ref{fig:idspectra} has a blue continuum with Balmer and He\,{\scriptsize II}\,$\lambda$4686 emission.  He\,{\scriptsize I}\,$\lambda$6678 is double-peaked in the average spectrum (Fig.~\ref{fig:avspectra}).  We obtained one light curve of fairly low quality of this system; it is shown in Fig.~\ref{fig:lightcurves}.  The short gap in the light curve was caused by poor conditions.

We took a total of 34 spectra of H$\alpha$103135 over two nights at ESO.  The Fourier transform has maximum power at $7.40 \times 10^{-2}\,\mathrm{mHz}$ (3.76~h).  It is also possible that the correct period corresponds to the second strongest aliases, at 3.22~h.  Assuming that the highest amplitude alias represents the orbital modulation, the period is $3.76\pm0.02\,\mathrm{h}$.  The trailed spectrum shows that the H$\alpha$ line profile is double-peaked at some orbital phases (Fig.~\ref{fig:trails}).  

\subsection{H$\alpha$103959}
H$\alpha$103959 has a spectrum displaying Balmer, He\,{\scriptsize II}, and very weak He\,{\scriptsize I} emission lines, and a blue continuum (see Fig.~\ref{fig:idspectra} and \ref{fig:avspectra}).  Our two short light curves of this system are shown in Fig.~\ref{fig:lightcurves}; run RP45 was shifted vertically by $-0.6$, and horizontally by $+3$~d.

The orbital period is most likely $3.785\pm0.005\,\mathrm{h}$, but aliases at 3.267~h and 4.498~h cannot be ruled out.  The trailed spectrum shows what is probably a double-peaked disc component, as well as a fainter, higher velocity component.

\subsection{H$\alpha$112921}
Weak Balmer emission lines are detected in the spectrum of H$\alpha$112921, superimposed on a blue continuum.  The H$\alpha$ line profile is double-peaked in some of the higher resolution spectra, and He\,{\scriptsize I} is weakly detected.  We do not have any high-speed photometry of this system.

The sampling of the time-resolved spectroscopy, taken at both SAAO and CTIO, is unfortunately far from optimal, with long gaps between different observations; for this reason there is some uncertainty in cycle count.  The strongest signal in the Fourier transform is at $7.538 \times 10^{-2}\,\mathrm{mHz}$ (corresponding to 3.685~h).  The next largest amplitude aliases are at 3.196~h and  4.352~h (see Fig.~\ref{fig:rvandfts}).  Assuming we have identified the correct alias, the orbital period is $3.6851\pm0.0004\,\mathrm{h}$.  The radial velocity amplitude is quite low (the fit gives $96.9\,\mathrm{km/s}$), but there is also a fainter, much higher velocity modulation visible in the trailed spectrum (Fig.~\ref{fig:trails}).

\subsection{H$\alpha$130559}
The spectrum of H$\alpha$130559 shows strong He\,{\scriptsize II}\,$\lambda$4686, and broad Balmer and He\,{\scriptsize I} lines (see Fig.~\ref{fig:idspectra} and \ref{fig:avspectra}).  We did not obtain any photometry of this system.

Time-resolved spectra of H$\alpha$130559 were taken on 6 and 8 March 2007.  The largest amplitude signal in the radial velocity curve of this system is at $7.072 \times 10^{-2}\,\mathrm{mHz}$ (3.928~h).  However, there is clearly a serious aliasing problem (see the Fourier transform in Fig.~\ref{fig:rvandfts}).  The two nearest aliases are at 4.273~h and 3.635~h.  Assuming that the peak marked in the Fourier transform represents the orbital modulation, the orbital period is $3.928\pm0.013\,\mathrm{h}$.  The radial velocity amplitude is large (the best fit has $244.4\,\mathrm{km/s}$).  The trailed spectrum in  Fig.~\ref{fig:trails} shows only an S-wave.

\subsection{H$\alpha$075648 and H$\alpha$122105}
\label{sec:075648and122105}
We obtained time-resolved spectroscopy of H$\alpha$075648 and H$\alpha$122105, but were not able to measure orbital periods from the data.  The spectra of both objects contain Balmer and He\,{\scriptsize I} lines in emission (Fig.~\ref{fig:idspectra} and Fig.~\ref{fig:avspectra}).  We have only a single short light curve of H$\alpha$075648, taken under poor conditions, but it does show the flickering expected in a mass transferring system.  The photometric variations in H$\alpha$122105 are faster and of slightly larger amplitude.  Neither system shows a photometric modulation that is readily recognizable as orbital.  A Fourier transform of the run RP31 light curve of H$\alpha$122105 shows that it contains a modulation with a period of roughly 920~s (this time-scale can be picked out in the light curve by eye).  The signal was not present on the next night, and was probably a QPO (CVs in which QPOs have been detected usually do not persistently display them, even when at the same brightness; e.g. \citealt{Warner04}; \citealt{PretoriusWarnerWoudt06}).  Horizontal and vertical shifts of $-1$~d and $-0.6$~mag were applied to the shorter light curve of H$\alpha$122105 (run RP34) displayed in Fig.~\ref{fig:lightcurves}.  The photometric and spectroscopic appearance of these two objects leave little doubt that they are both CVs.

We observed H$\alpha$075648 and H$\alpha$122105 for 7.3~h and 5.1~h, respectively, over 2 nights with EMMI, but failed to obtain plausible radial velocity curves.  Radial velocities were measured using cross correlation with template spectra, and also both a single Gaussian and a double Gaussian with a range of separations, for both the H$\alpha$ and He\,{\scriptsize I}\,$\lambda$5876 lines.  None of these techniques yielded sinusoidal radial velocity curves.  The results obtained are reminiscent of the radial velocity curve of the SW Sex star V380 Oph, displayed in fig.~13 of \cite{Rodriguez-GilSchmidtobreickGansicke07}, which shows a shorter time-scale variation superimposed on the sinusoidal orbital modulation.  If the same behaviour is displayed by H$\alpha$075648 and H$\alpha$122105, then we observed less than one orbital cycle in any one night.

\begin{figure}
 \includegraphics[width=84mm]{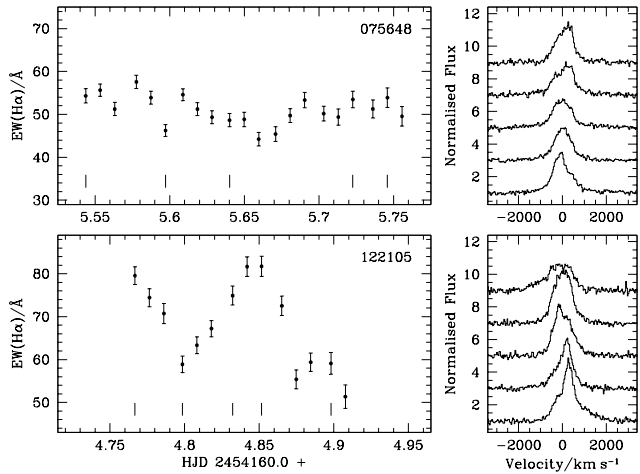}
 \caption{Left hand panels show the equivalent widths of the H$\alpha$ lines of H$\alpha$075648 and H$\alpha$122105 as a function of time for the longest spectroscopic runs on each of the systems.  The H$\alpha$ profiles of some spectra of H$\alpha$075648 (top) and H$\alpha$122105 (bottom) are shown on the right.  These spectra are normalized to the continuum and vertically offset.  Tick marks along the bottom of the EW curves indicate the spectra displayed in the right-hand panels; time increases upwards.  Velocity is relative to the rest wavelength of H$\alpha$.
}
 \label{fig:ews_profiles}
\end{figure}

The reason for the difficulty in finding radial velocities appears to be that the line profiles change dramatically on quite short time-scales, so that different velocity components are measured in different spectra.  H$\alpha$122105 also displays large variability in EW(H$\alpha$).  This is illustrated in Fig.~\ref{fig:ews_profiles}.  All standard techniques of measuring radial velocities assume that (at least) the line wings are constant in shape.  The failure of these techniques here are therefore not surprising.

EW flaring is observed in intermediate polars and SW Sex stars (e.g. \citealt{Morales-RuedaStillRoche96}; \citealt{SmithDhillonMarsh98}; \citealt{Rodriguez-GilMartinez-Pais02}).  
Strong He\,{\scriptsize II}\,$\lambda$4686 emission would support an interpretation of these two systems as magnetic CVs; however, it is not detected in either system, and, if it is present, it is significantly weaker than H$\beta$.
The very structured and variable line profiles, as well as the EW variations in H$\alpha$122105, may indicate SW Sex star behaviour, but additional observations, perhaps of higher time and/or spectral resolution, are required to determine the CV types of these two systems, and measure their periods.

\subsection{H$\alpha$163447}
In addition to the noisy discovery observation, we obtained a few higher resolution spectra, as well as a light curve, of H$\alpha$163447.  Spectra of the system display broad Balmer emission lines, and also He\,{\scriptsize I}\,$\lambda$5876 emission (see the average spectrum in Fig.~\ref{fig:avspectra}).  The light curve plotted in Fig.~\ref{fig:lightcurves} also shows the flickering typical of CVs.  We are confident in the classification of H$\alpha$163447 as a CV, but neither our time resolved spectroscopy nor photometry allows us to measure an orbital period.

\subsection{H$\alpha$115927 and H$\alpha$190039}
We obtained only identification spectra for H$\alpha$115927 and H$\alpha$190039, and both these spectra have low $S/N$.  The objects are classified as CV candidates, since their spectra show resolved Balmer emission lines, flat Balmer decrements, and possibly He\,{\scriptsize II}\,$\lambda$4686 emission.  However, better observations are needed to confirm the nature of these two objects.

\section{Distance limits and classification}
\label{sec:distances}
The semi-empirical donor sequence of \cite{Knigge06} predicts absolute magnitudes of CV secondaries as a function of orbital period, and provides a useful technique for constraining the distances of CVs.  We have used observed $K_S$ magnitudes (or lower limits on $K_S$), and the predicted absolute $K_S$ magnitudes\footnote{The CIT magnitudes of \cite{Knigge06} were transformed to the 2MASS system using the prescription of \cite{Carpenter01}, updated to reflect the final 2MASS data release (the updated colour transformations are available at http://www.astro.caltech.edu/$\sim$jmc/2mass/v3/transformations/).} of the secondaries in the systems for which we have period measurements, to find lower limits on their distances.  Since the new CVs are at low Galactic latitudes, it is not really appropriate to neglect extinction.  Therefore we took the total Galactic $N_H$ along the line of sight to each of our objects from the Galactic H\,{\scriptsize I} map of \cite{Kalberla05}, and estimated upper limits on $A_{K_S}$.  We  used relations between $N_H$ and $A_V$ and between $A_V$ and $A_{K_S}$ from \cite{PredehlSchmitt95} and \cite{CambresyBeichmanJarrett02}.  The resulting lower limits on distances are listed in Table~\ref{tab:summary}.

H$\alpha$092134 is in the period gap (the upper edge of the period gap is not very clearly defined, but \citealt{Knigge06} places it at at 3.18~h), although it is clearly accreting (see the light curve and spectra).  We used the predicted $M_{K_S}$ of a typical CV secondary at 3.18~h for the upper limit on its distance.  If the secondary in H$\alpha$092134 differs from a typical secondary in a CV at 3.18~h only in being of slightly lower mass, it would be fainter than we assume, and the distance limit we give would be too high.  On the other hand, the fact that 2MASS provides only a lower limit on $K_S$ takes our limit in the opposite direction.  2MASS photometry also gives only limits on $K_S$ for H$\alpha$073418, H$\alpha$092751, and H$\alpha$130559, implying that the lower limits on the distances of these three systems are conservative.  The distance limits assume that the predicted luminosities of the secondaries are appropriate for our systems (i.e. that the secondaries are unevolved).  This assumption is more likely to be valid for CVs at $P_{orb} \la 5 \mathrm{h}$.

The lower distance limits do not rule out the absolute magnitudes of quiescent dwarf novae (DNe) for any of these CVs, and therefore do not lead to firm classifications.

A blue continuum and weak emission lines characterise the spectra of many NLs.  The spectroscopic appearance of several of the new CVs indicate that they are very likely NLs, with H$\alpha$092751 and H$\alpha$112921 being the clearest examples (see Fig.~\ref{fig:idspectra}).

The different photographic $R$-band epochs, as well as the new observations presented here, provide good evidence of large amplitude variability for the following systems: H$\alpha$073418, H$\alpha$094409, H$\alpha$102442, and H$\alpha$163447.  These data are, however, not sufficient to distinguish between, e.g., DN and VY Scl star behaviour.

\section{Summary}
We have discovered 16 CV candidates by selecting emission line objects from the SHS, and obtained additional observations for 14 of these systems, confirming their CV nature.  Orbital periods were measured for 11 of the new CVs.  All of these are long-period systems, and most have orbital periods in the range 3 to 4 h.  The periods are listed in Table~\ref{tab:summary}.  Note that 6 of these periods were determined from aliased radial velocity curves, and may therefore only be correct to within $\sim 10$\% (the errors given in the table are for the strongest alias only).

The aim of this study is to construct a new CV sample with uniform selection criteria.  Although the bright flux limit implies that the sample is biased against intrinsically faint CVs, there is no explicit second selection cut that compounds this bias (but note that there is an effective $R-I$ cut, which will be discussed in Paper II).  It is also the largest H$\alpha$-selected CV sample constructed to date.

The blue, weak-lined spectra of H$\alpha$092751 and H$\alpha$112921 imply that they are probably both NLs.  Given the orbital periods of our CVs, it is likely that many of them are SW Sex stars.  Specifically, we have reason to suspect that H$\alpha$073418, H$\alpha$094409, H$\alpha$075648, and H$\alpha$122105 are SW Sex stars, but confirmation is needed for all of them.  H$\alpha$075648 and H$\alpha$122105 are unusual in displaying large variations in line profiles, which prevented us from measuring orbital periods.  H$\alpha$073418, H$\alpha$094409, H$\alpha$102442, and H$\alpha$163447 have been observed to display large amplitude variability.

The secondary star is detected in the spectrum of H$\alpha$074208.  We measure a spectral type of M$0\pm1$, and estimate that the secondary contributes between 40 and 80\% of the flux in the wavelength range from $\simeq5\,700$ to $\simeq7\,200\,\mathrm{\AA}$.

\section*{Acknowledgments}
MLP acknowledges financial support from the South African National Research Foundation and the University of Southampton.  We thank Tom Marsh for the use of his software package {\sc molly}, and Darragh O'Donoghue for helpful discussions.  Thanks also to the referee, Stuart Littlefair, for suggestions that have improved this paper.

\bsp

\label{lastpage}

\end{document}